\documentclass[journal]{IEEEtran}  
\usepackage{graphicx}
\usepackage{color, colortbl}
\usepackage{subfigure}
\usepackage{epsfig}
\usepackage{amsmath}
\usepackage{amsfonts}
\usepackage{amssymb}
\usepackage{cite}
\usepackage{float}

\begin{document}
\title{Antenna Selection in Polarization Reconfigurable MIMO (PR-MIMO) Systems
\thanks{Paul S. Oh and Sean S. Kwon are co-first authors with the equal contribution.}
\thanks{Part of this work was presented at IEEE Globecom 2015 \cite{Kwon_Molisch_GLOBECOM15}.}
}

\author{ Paul S. Oh$^\dagger$, Student Member, IEEE, Sean S. Kwon$^\dagger$, Member, IEEE and Andreas F. Molisch$^*$, Fellow, IEEE\\
$^\dagger$Department of Electrical Engineering, California State University Long Beach\\
$^*$Department of Electrical and Computer Engineering, University of Southern California\\
Email: {~}paul.oh@student.csulb.edu,{~~}sean.kwon@csulb.edu,{~~}molisch@usc.edu} \maketitle

\begin{abstract}
Adaptation of a wireless system to the polarization state of the propagation channel can improve reliability and throughput. This paper in particular considers polarization reconfigurable multiple-input multiple-output (PR-MIMO) systems, where both transmitter and receiver can change the (linear) polarization orientation at each element of their antenna arrays. We first introduce joint polarization pre-post coding to maximize bounds on the capacity and the maximum eigenvalue of the channel matrix. For this we first derive approximate closed form equations of optimal polarization vectors at one link end, and then use iterative joint polarization pre-post coding to pursue {\em joint} optimal polarization vectors at both link ends. Next we investigate the combination of PR-MIMO with hybrid antenna selection (PR-HS-MIMO), which can achieve a remarkable improvement of channel capacity. Two novel schemes of element-wise and global polarization reconfiguration are presented for PR-HS-MIMO. Comprehensive simulation results indicate that the proposed schemes provide 3 -- 5 dB signal-to-noise ratio (SNR) gain in PR-MIMO and concomitant improvements of channel capacity in PR-HS-MIMO.   
\end{abstract}

\IEEEpeerreviewmaketitle

\section{Introduction}

An electromagnetic wave that propagates from the transmitter (Tx) to the receiver (Rx) of a wireless communications system is characterized, {\em inter alia}, by its polarization, i.e., orientation of the field vector. Interaction of a wave (multipath component, MPC) with environmental objects may change the orientation, and different MPCs may experience different changes in the orientation. This effect can reduce performance, e.g., if the (fixed) polarization of the Rx antenna is mis-matched to the polarization of the arriving field. However, it can also be an advantage, since the different polarizations offer degrees of freedom that can be exploited for diversity and/or spatial multiplexing. Therefore, numerous investigations have been made into use of the polarization domain \cite{Kwon_Molisch_GLOBECOM15, Zhang_TCom20, Kwon_IGESSC20_SM, Zafari_Sari_TCom17, Nardelli_Ding_Cost_TWC19_MP_NOMA, Kwon_Oh_VTC20Fall, Hanzo_Pol_Hybrid_BF, qin2017compound, wolosinski20162, babakhani2016frequency, sun2016novel, cai2017continuously, liao2015polarization, Kwon14_GLOBECOM}. 
It has been demonstrated that utilizing the polarization domain may increase channel capacity and spectral efficiency; and improve symbol error rate (SER) \cite{Jootar06, White05, Erceg_Paulraj02, Kwon_Molisch_GLOBECOM15, Kwon_Stuber14_PDMA_TWC}. For this reason, impact of polarization on the wireless communication systems has been regarded as a promising research topic \cite{Kwon_Molisch_GLOBECOM15, Andrews01, Erceg04, Shafi_Molisch06, Landmann07, Kwon_Stuber11_TVT, Kwon_Oh_VTC20Fall, Kwon_IGESSC20_SM}.

In particular, several recent research works present the benefit of utilizing the polarization domain  in recently proposed communication schemes including, but not limited to, MIMO spatial multiplexing \cite{Kwon_Molisch_GLOBECOM15}; spatial modulation (SM) \cite{Zhang_TCom20, Kwon_IGESSC20_SM, Zafari_Sari_TCom17}; non-orthogonal multiple access (NOMA) \cite{Nardelli_Ding_Cost_TWC19_MP_NOMA}; and beamforming \cite{Kwon_Oh_VTC20Fall, Hanzo_Pol_Hybrid_BF}. It is validated that the deployment of dual-polarized antennas can not only increase channel capacity \cite{Andrews01, Kwon14_GLOBECOM, Shafi_Molisch06, Kwon_Thesis}; but also improve SER \cite{Kwon_Oh_VTC20Fall, Jootar06, Erceg_Paulraj02, White05, Kwon_Stuber14_PDMA_TWC}. Hybrid beamforming can adopt dual polarization and the associated codebook design to improve the system performance \cite{Hanzo_Pol_Hybrid_BF, Kim_Love10}. Although polarization multiplexing without spatial diversity is promising \cite{Kwon_Stuber14_PDMA_TWC}, polarization diversity can be combined with spatial diversity to further improve the performance of wireless communication systems \cite{Jootar06, White05, Erceg_Paulraj02, Kwon_Molisch_GLOBECOM15}.

Various other aspects of polarization in MIMO systems have been investigated as well. Ref. \cite{White05} showed that space-time block coding (STBC) with single polarization outperforms STBC with dual polarization in Rayleigh and Ricean fading channels. A MIMO system with dual-polarized antenna elements can have lower spatial diversity but higher spatial multiplexing gain than a conventional MIMO system with single-polarized antennas, particularly, in Ricean fading channels with high $K$-factor \cite{Erceg_Paulraj02}. It is noteworthy that the extent of benefit from dual-polarized antennas depends on the associated schemes to exploit the characteristics of polarized wireless channel \cite{Jootar06, White05, Erceg_Paulraj02, Kwon_Molisch_GLOBECOM15, Kwon_Oh_VTC20Fall}. Various channel sounding campaigns and channel models provide insights into the characteristics of wireless channel polarization \cite{Hong_Pol_BF, Shafi_Molisch06, Landmann07, Erceg04, Soma02, Paulraj00, Kwon_Stuber11_TVT, Kwon_Stuber13_TVT, Kwon_Stuber13_TVTS}.

Most of the aforementioned papers
\cite{Erceg_Paulraj02,Kwon_Thesis,Kwon14_GLOBECOM,Kwon_Stuber14_PDMA_TWC}, 
assume that the
Tx and Rx have co-located dual-polarized antennas, such that two antenna ports are available for each antenna element at a distinct spatial location. Compared to the case where the same number of antenna elements with
only a single polarization is available, this leads to an increase in diversity and capacity, although the gain depends significantly on the XPD \cite{Jootar06, Kwon_Oh_VTC20Fall}. However, this also leads to doubling the number of antenna ports,
with the consequent increase in the number of RF (up/down-conversion) chains and increased baseband processing. 
To mitigate the resulting increase in energy consumption and complexity, this paper considers polarization reconfigurable multi-input multi-output (PR-MIMO) systems, where each antenna element has only a single port, but whose polarization orientation can be adapted through use of switches and parasitic elements \cite{cai2017continuously, sun2016novel, babakhani2016frequency, Landon_Furse07, Karabey_Jakoby13}. The performance analysis and transceiver schemes in this paper are based on these types of antennas. 
%
By optimally employing the polarization reconfigurable antennas in conjunction with MIMO spatial multiplexing, this paper will show that the system performance is substantially enhanced compared to that of a conventional scheme with the same number of antenna ports. Throughout the paper, we will assume perfect knowledge of channel state information (CSI), including polarization state information, at both Tx and Rx. The methods for acquiring such CSI have been amply discussed in the literature.

Another important method for improving the cost/performance tradeoff in spatial multiplexing systems is hybrid antenna selection (HS) \cite{Molisch_ReducedComplexity, Win_MZ_Winter_HybridSelection, Hybrid_Selection, Gore_AntennaSelection,Receiver_antennaselection}. HS schemes can reduce hardware complexity by lowering the number of RF chains from $N_t$ to $L_t$. They are adopted in current communication standards including IEEE 802.11 \cite{zhang2006applying} and LTE \cite{mehta2012antenna}.  
Further, 5th generation (5G) and beyond-5G base stations (BSs), called next-generation Node B (gNB), consider deploying a large number of antenna elements but with the limited number of antenna ports in each antenna panel as described in \cite{Roh_5GBeamforming_ComMag14, Nam_FD-MIMO, Nam_FD-MIMO_Feasibility_JSAC17, Kwon_Oh_VTC20Fall}, which can be exploited by HS, possibly in combination with (hybrid) beamforming \cite{ratnam2018hybrid}.  
 
On the other hand, polarization diversity is not taken into account in the majority of previous research works on antenna selection. Although there are previous reports that consider polarization diversity with antenna selection, they consider fixed antenna polarization such as dual-polarized antennas in\cite{Anreddy_Vikram_Ingram_Ann} or tri-polarized antennas in \cite{Aamir_MP_AntennaSelection}. To the best of our knowledge, the current paper is the first to exploit polarization reconfigurable HS - MIMO spatial multiplexing (PR-HS-MIMO), which significantly outperforms that of the conventional HS-MIMO systems with comparable complexity.  

The primary contributions of this paper can be summarized as follows:
\begin{itemize}
  \item we introduce and characterize the hybrid antenna selection at the Tx with polarization reconfigurable antennas on both link ends;
  \item we provide a closed-form approximation of the optimal
      Tx/Rx-polarization vectors in the PR-MIMO system at one link end to achieve channel capacity beyond that of a conventional MIMO communication system (with the same number of ports). We also propose iterative joint polarization pre-post coding for joint optimal polarization vectors;
  \item we propose two novel schemes of PR-HS-MIMO with polarization reconfigurable antenna elements, and provide analysis of the system performance in terms of channel capacity and the distribution of channel gain;      
  \item we perform a statistical analysis of the effect of polarization reconfigurable antennas on the distribution of channel gain;
  \item we validate the significant improvement of the system performance in terms of channel capacity caused by the proposed PR-MIMO and PR-HS-MIMO schemes.
\end{itemize}

The remainder of this paper is organized as follows. Section \ref{sec:PR-MIMO_System}
presents the PR-MIMO system model focusing on polarization reconfigurable antennas. In
Section \ref{sec:Pol_Pre-Post_Coding}, the iterative joint polarization pre-post coding
scheme utilizing optimal Tx/Rx-polarization vectors is proposed and described
in detail. PR-HS-MIMO is presented in Section \ref{sec:AntSelectionNPolarization}; the statistical analysis of the polarization reconfigurable channel is described in Section \ref{sec:Stat of Polarization}.
Extensive simulation results demonstrate significant improvement of the system performance in terms of channel capacity and statistics of channel gain in Section \ref{sec:Results}. Finally, Section \ref{sec:Conclusion} concludes the paper.

\section{System Model}
\label{sec:PR-MIMO_System}

\begin{figure}[ht]
  \centering
  \includegraphics[width=0.49\textwidth]{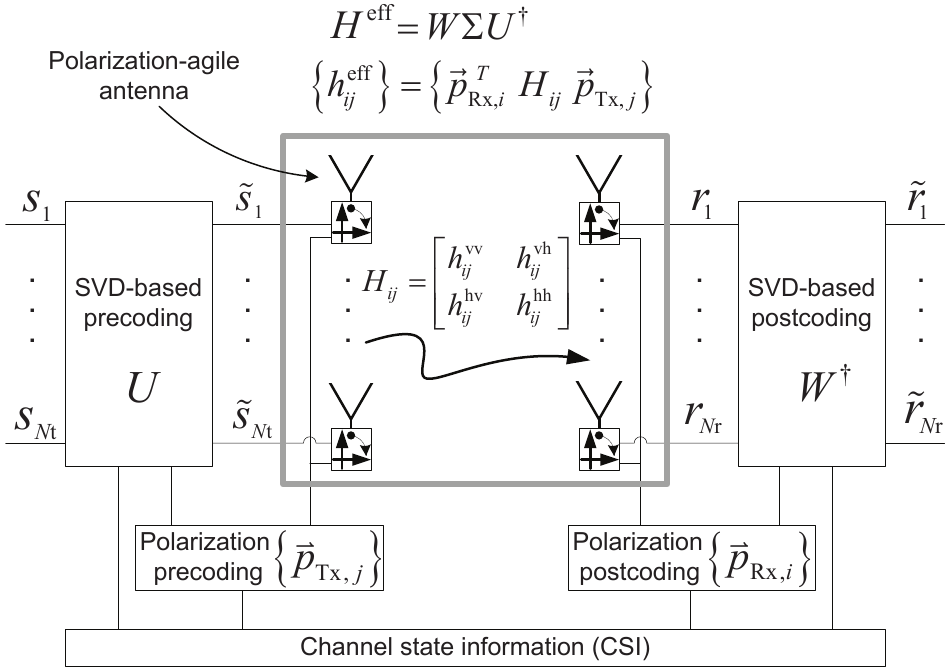}  
  \caption{PR-MIMO system with polarization reconfigurable antennas.}
  \label{fig:PR-MIMO_system}
\end{figure}

The fundamental block diagram of the PR-MIMO system is shown in Fig. \ref{fig:PR-MIMO_system}, where the Tx and the Rx have $N_{\rm t}$ and $N_{\rm r}$ antenna elements, respectively. Each antenna element is polarization reconfigurable with the polarization vector $\vec{p}_{{\rm Tx},j}$ and $\vec{p}_{{\rm Rx},i}$ in which $j \in \{1,...,N_{\rm t}\}$; $i \in
\{1,...,N_{\rm r}\}$. The polarization vector is adjusted according to the channel
state information (CSI); perfect CSI at all antenna elements is assumed to be available at the Rx (CSIR) as well as the Tx (CSIT). It is noteworthy that the CSI is available for both orthogonal polarization directions; it can be obtained through training schemes similar to those employed in antenna selection systems
(see, e.g., \cite{Molisch_and_Win_2004}). The impact of imperfect CSI is outside the scope of this paper and will be analyzed in future work.

In addition to the polarization precoding/postcoding, the PR-MIMO system contains precoding/postcoding that allows optimum exploitation of the spatial degree of freedom; it is well-known from conventional (non-polarization reconfigurable) spatial multiplexing systems with perfect CSI \cite{Molisch_2010_book} that
linear precoding/postcoding based on the (SVD) of the effective channel impulse matrix maximizes sum-rate. It is intuitive that exploiting polarization reconfigurable antenna elements with polarization
precoding/postcoding - on top of the standard SVD-based spatial
precoding/postcoding - can achieve higher or equally good capacity than single-polarization or
fixed-polarization antennas (with the same number of data streams or RF
up/down-conversion chains). Mathematical derivations for this intuition are presented below.

The effective channel impulse response matrix in Fig.~\ref{fig:PR-MIMO_system}
can be expressed as
\begin{equation}
 \label{eq:H_eff}
 H^{\rm eff} =
     \begin{bmatrix}
       \vec{p}_{{\rm Rx},1}^{~T} H_{11} \vec{p}_{{\rm Tx},1}
       & \ldots
       & \vec{p}_{{\rm Rx},1}^{~T} H_{1 N_{\rm t}} \vec{p}_{{\rm Tx},N_{\rm t}} \\
       \vdots & \ddots & \vdots \\
       \vec{p}_{{\rm Rx},N_{\rm r}}^{~T} H_{N_{\rm r} 1} \vec{p}_{{\rm Tx},1}
       & \ldots
       & \vec{p}_{{\rm Rx},N_{\rm r}}^{~T} H_{N_{\rm r} N_{\rm t}} \vec{p}_{{\rm Tx},N_{\rm t}}
     \end{bmatrix},
\end{equation}
where the operation $(\cdot)^T$ is the transpose of a given vector or matrix,
and the dimension of $H^{\rm eff}$ is $N_{\rm r} \times N_{\rm t}$.    Further, we denominate $H_{ij}$ as ``$\emph{polarization-basis matrix}$'' which is
\begin{equation}
 \label{eq:Pol_basis_matrix}
 H_{ij} =
     \begin{bmatrix}
       h_{ij}^{\rm vv} & h_{ij}^{\rm vh} \\
       h_{ij}^{\rm hv} & h_{ij}^{\rm hh} \\
     \end{bmatrix},
\end{equation}
where $h_{ij}^{\rm xy}$ with ${\rm x \in \{v,h\} }$; ${\rm y \in \{v,h\} }$ is
the XY-channel impulse response from the Y-polarization Tx antenna to the
X-polarization Rx antenna. For instance, $h_{ij}^{\rm hv}$ is the channel
impulse response from the vertically polarized (V-Pol) Tx antenna to
the horizontally polarized (H-Pol) Rx antenna. We assume here a flat-fading channel, such that the $h_{ij}^{\rm xy}$ are complex scalars.  Lastly,
$\vec{p}_{{\rm Tx},j}$ and $\vec{p}_{{\rm Rx},i}$ are, respectively, the
Tx-polarization vector at the $j$th Tx antenna and the Rx-polarization vector
at the $i$th Rx antenna, and they are expressed as
\setlength{\arraycolsep}{0.14em}
\begin{eqnarray}
 \label{eq:Tx-Pol_vector}
 \vec{p}_{{\rm Tx},j} &=&   \begin{bmatrix}
                              p_{{\rm Tx},j}^{\rm v} \\
                              p_{{\rm Tx},j}^{\rm h} \\
                            \end{bmatrix}
                        =   \begin{bmatrix}
                              \cos \theta_j \\
                              \sin \theta_j \\
                            \end{bmatrix}, \\
 \vec{p}_{{\rm Rx},i} &=&   \begin{bmatrix}
                              p_{{\rm Rx},i}^{\rm v} \\
                              p_{{\rm Rx},i}^{\rm h} \\
                            \end{bmatrix}
                        =   \begin{bmatrix}
                              \cos \theta_i \\
                              \sin \theta_i \\
                            \end{bmatrix}. \label{eq:Rx-Pol_vector}
\end{eqnarray}
\setlength{\arraycolsep}{5pt}Here, we call the angles $\theta_j$ and $\theta_i$
Tx- and Rx-polarization angles, respectively.   It is worth mentioning that Tx-
and Rx-polarization vectors are unit vectors so that the overall signal power
is preserved after polarization precoding and postcoding.

\section{Polarization Pre-post Coding at the Polarization reconfigurable Antenna}
\label{sec:Pol_Pre-Post_Coding}


\subsection{Polarization Precoding and Postcoding with Optimal Tx- and Rx-polarization}
\label{sec:Pol_Precoding}

The Tx and the Rx can utilize SVD-based precoding and postcoding under the
assumption of full CSIT and CSIR. The combination of SVD-based
precoding and postcoding achieves the MIMO channel capacity for a given channel
matrix via constructing parallel channels
\cite{Molisch_2010_book}.   On the other hand, the MIMO communication system
with polarization reconfigurable antennas in Fig.~\ref{fig:PR-MIMO_system} can tune the effective channel impulse response matrix, $H^{\rm eff}$ in
(\ref{eq:H_eff}) by either polarization precoding at the Tx or polarization
postcoding at the Rx; or joint polarization pre-post coding at both ends.   We
first focus on polarization precoding with fixed $\vec{p}_{{\rm Rx},i}$ in this section.

The effective channel impulse response matrix can be decomposed by SVD as
\begin{equation}
 H^{\rm eff} = W \Sigma U^{\dag},
\end{equation}
where $\Sigma$ is a diagonal matrix containing singular values, and $W$ and
$U^{\dag}$ are unitary matrices composed of the left and right singular
vectors, respectively \cite{Molisch_2010_book}.   In this paper, $(\cdot)^\dag$
is the Hermitian transpose operation.   The channel capacity with SVD-based
precoding and postcoding is
\begin{equation}
 \label{eq:C_sum}
 C = \sum _{l=1}^{ R_{H^{\rm eff}} } \log_2 \left( 1 + \frac{P_l}{{\sigma_{\rm n}}^2} {\sigma_l}^2 \right),
\end{equation}
where $R_{H^{\rm eff}}$ is the rank of the matrix $H^{\rm eff}$, and $P_l$ is
the power allocated to the $l$th eigenmode.   Further, $\sigma_l$ is the $l$th
singular value of the effective channel impulse response matrix $H^{\rm eff}$,
and ${\sigma_{\rm n}}^2$ is the noise power.   Capacity maximization is achieved by a power allocation $P_l$ that satisfies
waterfilling conditions
\begin{equation}
 P_ = \max \left( 0, \epsilon - \frac{ {\sigma_{\rm n}}^2 }{ {\sigma_l}^2 } \right) {\rm~s.t.~}
 P = \sum _{l=1}^{R_{ H^{\rm eff} }} P_l {~}, \label{eq:Power_Constraint}
\end{equation}
i.e., the threshold $\epsilon$ is determined by the constraint of the total
transmitted power P.
We assume the total power is independent of the number of antennas.

Following \cite{Tse_and_Visvanath_2005_book}, we use Jensen's inequality to obtain
\setlength{\arraycolsep}{0.14em}
\begin{eqnarray}
 \label{eq:Capacity_Jensen_Ineq}
 C &\leq& R_{H^{\rm eff}} {~} \log_2 \left( 1 + \frac{1}{ R_{H^{\rm eff}} }
   \sum _{l=1}^{ R_{H^{\rm eff}} } \frac{P_l}{{\sigma_{\rm n}}^2} {\sigma_l}^2 \right) \\
 &=& R_{H^{\rm eff}} {~} \log_2 \left( \frac{1}{ R_{H^{\rm eff}} }
   \frac{\epsilon}{ {\sigma_{\rm n}}^2 }  \sum _{l=1}^{ R_{H^{\rm eff}} } {\sigma_l}^2 \right),
 \label{eq:Capacity}
\end{eqnarray}
\setlength{\arraycolsep}{5pt}where high SNR is assumed;
all $P_l$ is greater than zero, i.e., $P_l = \epsilon - {\sigma_n}^2 /
{\sigma_l}^2 > 0$ in (\ref{eq:Power_Constraint}).   Further, ${\sigma_l}$ is
the singular value of $H^{\rm eff}$; therefore
\cite{Linear_Agebra_Strang}, \setlength{\arraycolsep}{0.14em}
\begin{eqnarray}
 \label{eq:Sum_Squared_SV}
 \sum _{l=1}^{ R_{H^{\rm eff}} } {\sigma_l}^2 &=&
   {\rm Tr} \left( H^{\rm eff} (H^{\rm eff})^\dag \right)
     = \sum _{n,m} \left| h_{nm}^{\rm eff} \right|^2.
\end{eqnarray}
\setlength{\arraycolsep}{5pt}This is the sum of squared envelopes of all
channel impulse response elements in $H^{\rm eff}$. This quantity is not only important for the upper bound of channel capacity, but will also play an important role in the polarization pre-post coding and antenna selection in PR-HS-MIMO systems. 

It is worth emphasizing that each element of
$H^{\rm eff}$ is affected by the Tx- and Rx-polarization as implied in
(\ref{eq:H_eff}).   Hence, the polarization vectors at polarization reconfigurable antenna elements impact constructively or destructively the MIMO channel capacity itself, even though SVD-based precoding/postcoding reaches the MIMO channel capacity for the given full CSIR/CSIT. 
The $j$th Tx polarization-agile antenna affects the $j$th column in
(\ref{eq:H_eff}); thus, the sum of squared singular values in
(\ref{eq:Sum_Squared_SV}) can be written as \setlength{\arraycolsep}{0.14em}
\begin{eqnarray}
 \sum _{l=1}^{ R_{H^{\rm eff}} } {\sigma_l}^2 &=&
   \sum _{j=1}^{N_{\rm t}} \sum _{i=1}^{N_{\rm r}}
   \vec{p}_{{\rm Tx},j}^{~T}
   \left( H_{ij}^{\dag} \vec{p}_{{\rm Rx},i} \vec{p}_{{\rm Rx},i}^{~T} H_{ij} \right)
   \vec{p}_{{\rm Tx},j} \nonumber\\
 &=& \sum _{j=1}^{N_{\rm t}} \vec{p}_{{\rm Tx},j}^{~T}
     H_{{\rm Tx},j}^{\rm PD}
     \vec{p}_{{\rm Tx},j} {~}, \label{eq:Sum_Sq_Singular}
\end{eqnarray}
\setlength{\arraycolsep}{5pt}where the ``\emph{Tx-polarization-determinant
matrix}'' for the $j$th Tx polarization-agile antenna, $H_{{\rm Tx},j}^{\rm
PD}$, is defined as \setlength{\arraycolsep}{0.14em}
\begin{eqnarray}
 \label{eq:Tx-Pol_Det}
 H_{{\rm Tx},j}^{\rm PD} &\triangleq& \sum _{i=1}^{N_{\rm r}}
   H_{ij}^\dag \vec{p}_{{\rm Rx},i} \vec{p}_{{\rm Rx},i}^{~T} H_{ij} {~}.
\end{eqnarray}
\setlength{\arraycolsep}{5pt}The Tx-polarization vector at each Tx
polarization-agile antenna is independent of those at other Tx-polarization
antennas; therefore, the optimal Tx-polarization vector at the $j$th Tx
polarization-agile antenna, $\vec{p}_{{\rm Tx},j}$, is the one which maximizes
$\vec{p}_{{\rm Tx},j}^{~T} H_{{\rm Tx},j}^{\rm PD} \vec{p}_{{\rm Tx},j}$ in
(\ref{eq:Sum_Sq_Singular}).

\begin{figure}[ht]
  \centering
  \includegraphics[width=0.45\textwidth]{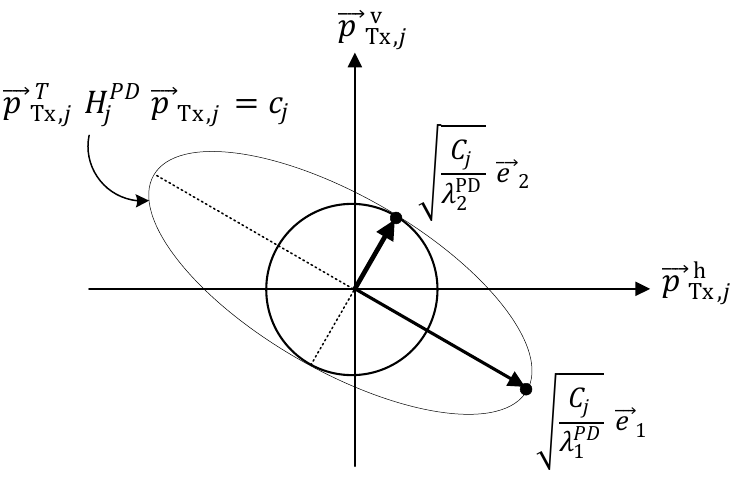}  
  \caption{Polarization-determinant ellipse and polarization-vector unit circle.}
  \label{fig:Pol_Det_Ellipse}
\end{figure}

From the viewpoint of the linear-algebraic approach, $\vec{p}_{{\rm Tx},j}^{~T}
H_{{\rm Tx},j}^{\rm PD} \vec{p}_{{\rm Tx},j}$ is positive semi-definite; thus,
the equation, $\vec{p}_{{\rm Tx},j}^{~T} {~} H_{{\rm Tx},j}^{\rm PD} {~}
\vec{p}_{{\rm Tx},j} = c_j$ corresponds to the ellipse as portrayed in
Fig.~\ref{fig:Pol_Det_Ellipse}, where $x$ and $y$ coordinates correspond to the
elements of $\vec{p}_{{\rm Tx},j}$, i.e., $p_{{\rm Tx},j}^{\rm v}$ and $p_{{\rm
Tx},j}^{\rm h}$, respectively \cite{Linear_Agebra_Strang}. Geometrically, the
principal axes of the ellipse are along eigenvectors of the matrix $H_{{\rm
Tx},j}^{\rm PD}$, $\overrightarrow{e}_1$ and $\overrightarrow{e}_2$, and the
distances from the origin to the ellipse along the principal axes are
$\sqrt{c_j/\lambda_1^{\rm PD}}$ and $\sqrt{c_j/\lambda_2^{\rm PD}}$, where
$\lambda_1^{\rm PD}$ and $\lambda_2^{\rm PD}$ are the eigenvalues of $H_{{\rm
Tx},j}^{\rm PD}$.

Our objective at this stage is to estimate the optimal Tx-polarization vector
at each Tx polarization-agile antenna element, $\vec{p}_{{\rm Tx},j} \in j = \{
1, \cdots, N_{\rm t} \}$, which maximizes the column-sum of element-wise
squared envelopes in $H^{\rm eff}$, i.e., $\vec{p}_{{\rm Tx},j}^{~T} H_{{\rm
Tx},j}^{\rm PD} \vec{p}_{{\rm Tx},j}$.   On the other hand, the Tx-polarization
vector $\vec{p}_{{\rm Tx},j}$ is on the unit circle as presented in
(\ref{eq:Tx-Pol_vector}); therefore, the ellipse must have, at least, one
intersection or contact point with the unit circle; whereas, at the same time
it must make $\vec{p}_{{\rm Tx},j}^{~T} H_{{\rm Tx},j}^{\rm PD} \vec{p}_{{\rm
Tx},j} = c_j$ as large as it can.   Hence, the optimal Tx-polarization vector
$\vec{p}_{{\rm Tx},j}^{\rm opt}$ and corresponding optimal Tx-polarization
angle $\theta_j^{\rm opt}$ described in (\ref{eq:Tx-Pol_vector}) are as
\begin{eqnarray}
 \label{eq:Opt_Tx_Pol_vector}
 \vec{p}_{{\rm Tx},j}^{\rm ~opt} &=& \arg \max \limits_{ \vec{p}_{{\rm Tx},j} }{~}
   \vec{p}_{{\rm Tx},j}^{~T} H_{{\rm Tx},j}^{\rm PD} \vec{p}_{{\rm Tx},j}
     = {~} \vec{e}_2, \\
 \theta_j^{\rm opt} &=& \arctan(\vec{e_2}). \label{eq:Opt_Tx_Pol_angle}
\end{eqnarray}
Notice that $\vec{e}_2$ is the eigenvector corresponding to $\lambda_2$, which
is the maximum eigenvalue of the Tx-polarization-determinant matrix, $H_{{\rm
Tx},j}^{\rm PD}$.   In this manner, each Tx polarization-agile antenna element
can perform polarization precoding with the optimal Tx-polarization vector.

In a completely analogous manner, the optimal RX-polarization vector can be
derived; we just now have to employ the row-sum (instead of the column-sum) of
element-wise squared envelopes in $H^{\rm eff}$.  The optimum polarization
vector can be shown to be

\begin{eqnarray}
 \label{eq:Opt_Rx_Pol_vector}
 \vec{p}_{{\rm Rx},i}^{\rm ~opt} &=& \arg \max \limits_{ \vec{p}_{{\rm Rx},i} }{~}
   \vec{p}_{{\rm Rx},i}^{~T} H_{{\rm Rx},i}^{\rm PD} \vec{p}_{{\rm Rx},i}
     = {~} \vec{e}_2, \\
 \theta_i^{\rm opt} &=& \arctan(\vec{e_2}), \label{eq:Opt_Rx_Pol_angle}
\end{eqnarray}
\setlength{\arraycolsep}{5pt}where the ``\emph{Rx-polarization-determinant
matrix}'' for the $i$th Rx polarization-agile antenna, $H_{{\rm Rx},i}^{\rm
PD}$, is defined as \setlength{\arraycolsep}{0.14em}
\begin{eqnarray}
 \label{eq:Rx-Pol_Det}
 H_{{\rm Rx},i}^{\rm PD} &\triangleq& \sum _{j=1}^{N_{\rm t}}
   H_{ij} \vec{p}_{{\rm Tx},j} \vec{p}_{{\rm Tx},j}^{~T} H_{ij}^\dag {~}.
\end{eqnarray}
\setlength{\arraycolsep}{5pt}

\subsection{Joint Polarization Pre-post Coding for Polarization Matching}
\label{sec:Joint_Pol_Pre-Post_Coding}

The Tx-polarization-determinant matrix at the $j$th Tx polarization-agile antenna,
$H_{{\rm Tx},j}^{\rm PD}$, depends on the Rx-polarization vectors
$\vec{p}_{{\rm Rx},i}$ as shown in (\ref{eq:Tx-Pol_Det}), and vice versa in
(\ref{eq:Rx-Pol_Det}).   Further, Tx- and Rx-polarization mismatching will
deteriorate the system performance in terms of the channel capacity in this
paper.   For those reasons, joint polarization pre-post coding is required to
maximize PR-MIMO channel capacity.

Joint optimization of the pre-post coding is difficult to obtain in
closed form (and also difficult to implement); therefore, we propose an iterative approach
where one iteration is a sequential loop of polarization precoding; then
polarization postcoding.  In the $q$th iteration stage, $\vec{p}^{{\rm
~opt},(q)}_{{\rm Tx},j}$ is updated to $\vec{p}^{{\rm ~opt},(q+1)}_{{\rm
Tx},j}$ based on $\vec{p}^{{\rm ~opt},(q)}_{{\rm Rx},i}$ according to
(\ref{eq:Tx-Pol_Det})~--~(\ref{eq:Opt_Tx_Pol_angle}).  Then, in turn,
$\vec{p}^{{\rm ~opt},(q)}_{{\rm Rx},i}$ is updated to $\vec{p}^{{\rm
~opt},(q+1)}_{{\rm Rx},i}$ based on the updated transmit polarization vector, $\vec{p}^{{\rm ~opt},(q+1)}_{{\rm
Tx},j}$ following (\ref{eq:Opt_Rx_Pol_vector})~--~(\ref{eq:Rx-Pol_Det}). If initial values of Rx and Tx polarization vectors are
\begin{eqnarray}
 \label{eq:initial condition}
 \vec{p}_{{\rm Tx},j}^{\ (0)} &=&   \begin{bmatrix}
                              p_{{\rm Tx},j}^{\rm v} \\
                              p_{{\rm Tx},j}^{\rm h} \\
                            \end{bmatrix}
                        =   \begin{bmatrix}
                              1 \\
                              0 \\
                            \end{bmatrix}, \\
 \vec{p}_{{\rm Rx},i}^{\ (0)} &=&   \begin{bmatrix}
                              p_{{\rm Rx},i}^{\rm v} \\
                              p_{{\rm Rx},i}^{\rm h} \\
                            \end{bmatrix}
                        =   \begin{bmatrix}
                              1 \\
                              0 \\
                            \end{bmatrix},
\end{eqnarray}
the iterative joint optimization iteration is expressed as
\begin{eqnarray}
 \label{eq:Txpoljoint}
 H_{{\rm Tx},j}^{{\rm PD},(q+1)} &=& \sum _{i=1}^{N_{\rm r}}
   H_{ij}^\dag \vec{p}_{{\rm Rx},i}^{~(q)} \vec{p}_{{\rm Rx},i}^{~T(q)} H_{ij} {~}, \\
 \vec{p}_{{\rm Tx},j}^{{\rm ~opt} ,(q+1)} &=& \arg \max \limits_{ \vec{p}_{{\rm Tx},j} }{~}
   \vec{p}_{{\rm Tx},j}^{~T} H_{{\rm Tx},j}^{{\rm PD},(q+1)} \vec{p}_{{\rm Tx},j} {~}, 
  \label{eq:Opt_TxDeterminant}
\end{eqnarray}
\begin{eqnarray}
 \label{eq:Rxpoljoint}
 H_{{\rm Rx},i}^{{\rm PD},(q+1)} &=& \sum _{j=1}^{N_{\rm t}}
   H_{ij}^\dag \vec{p}_{{\rm Tx},j}^{~(q+1)} \vec{p}_{{\rm Tx},j}^{~T(q+1)} H_{ij} {~}, \\ 
  \vec{p}_{{\rm Rx},i}^{{\rm ~opt} ,(q+1)} &=& \arg \max \limits_{ \vec{p}_{{\rm Rx},i} }{~}
   \vec{p}_{{\rm Rx},i}^{~T} H_{{\rm Rx},i}^{{\rm PD},(q+1)} \vec{p}_{{\rm Rx},i} {~},\\ 
  \label{eq:Opt_RxDeterminant}
 \label{eq:FinalIteration}
 h^{\rm eff}_{\rm opt,ij}&=&(\vec{p}_{{\rm Rx},i}^{{\rm ~opt} ,(q+1)})^{T}H_{ij}\vec{p}_{{\rm Tx},j}^{{\rm ~opt} ,(q+1)}~,
\end{eqnarray}  
where q is the iteration index.
Note that while each step increases the capacity, the procedure is not guaranteed to
reach the {\em global} optimum. However, we will see in Section
\ref{sec:Results} that the capacity resulting from the iterative joint pre-post
coding is in a close agreement with the one achieved by brute-force numerical
search over all pre-post coding vectors.

\section{Antenna Selection with Reconfigurable Polarization}
 \label{sec:AntSelectionNPolarization}
\begin{figure*}[ht]
  \centering
  \includegraphics[width=0.90 \textwidth, height=2.5 in]{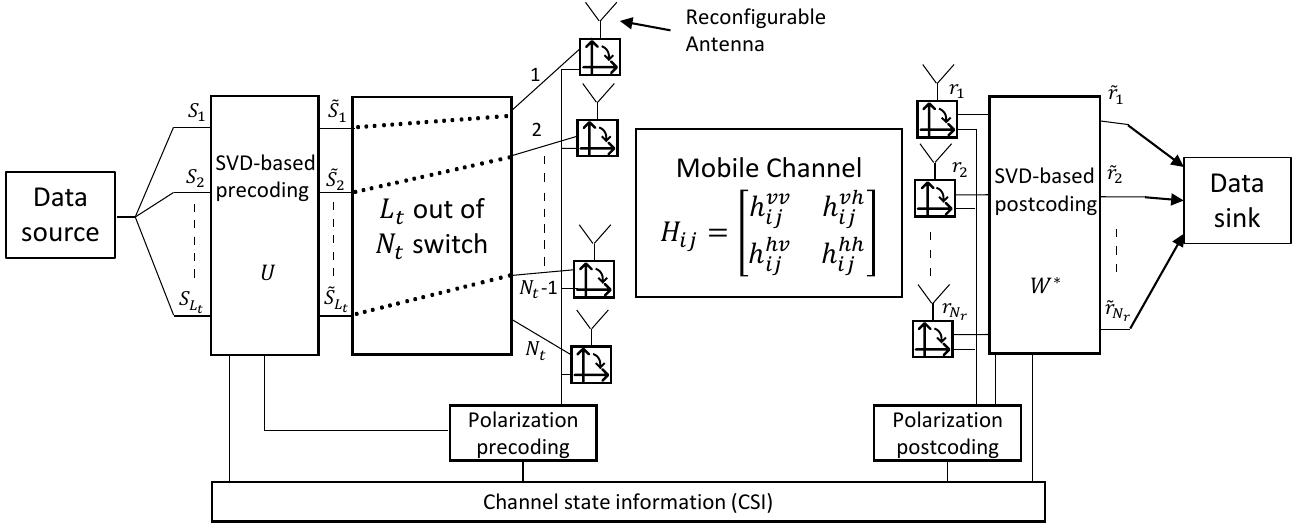}
  \caption{System model of polarization reconfigurable hybrid antenna selection in MIMO (PR-HS-MIMO) communications}
  \label{fig:AntSelePMIMOSystemModel}
\end{figure*}
This section delineates the PR-HS-MIMO communication system that performs spatial multiplexing with the selected antenna elements as depicted in Fig. \ref{fig:AntSelePMIMOSystemModel}. The Tx selects $L_t$ out of $N_t$ polarization reconfigurable Tx antenna elements, and performs spatial multiplexing based on SVD-based precoding and postcoding. Each antenna element is polarization reconfigurable and adjusted to an optimal polarization angle based on the joint polarization pre-post coding proposed and described in Section \ref{sec:Pol_Pre-Post_Coding}. In the same fashion as the PR-MIMO system, this section assumes perfect CSI. That is, both the Tx and Rx have perfect knowledge of the polarization-basis matrix in (\ref{eq:H_eff}) -- (\ref{eq:Pol_basis_matrix}).
 
The partial channel impulse response matrix that represents the channels between selected polarization reconfigurable Tx antenna elements and Rx antenna elements is defined as $\widetilde{H}^{\rm eff}_k$, where the index $k \in \big\{1, 2, ..., \binom{N_t}{L_t} \big\}$ stands for the $k$-th subset of $L_t$ antenna selection. Further, the column index matrix $C_{\rm Index}$ indicates the list of all possible combination of selected Tx antenna elements. That is,
\begin{equation}
 \label{eq:Column_Idx_matrix}
 C_{\rm Index} = 
 \begin{bmatrix}
 	\alpha_{11}
 	&\ldots
 	&\alpha_{1L_t} \\
 	&\vdots \\
 	\alpha_{\binom{N_t}{L_t}1} 
 	&\ldots
 	&\alpha_{\binom{N_t}{L_t}L_t}
 \end{bmatrix},
\end{equation}
where each row represents a subset of polarization reconfigurable Tx antenna indices corresponding to the selected antenna elements.

Each $\widetilde{H}^{\rm eff}_k$ consists of $L_t$ selected columns of $H^{\rm eff}$, associated with the selected polarization reconfigurable Tx antenna elements, and is described as
\begin{equation}\setlength{\arraycolsep}{0.14em}
 \label{eq:AllpossibleHeff}
 \begin{aligned}
 &&\widetilde{H}_k^{\rm eff} ={~~~~~~~~~~~~~~~~~~~~~~~~~~~~~~~~~~~~~~~~~~~~~~~~~~~~~~~~~~} \\
{}&&{}\begin{bmatrix}
       \vec{p}_{{\rm Rx},1}^{~T} H_{1 \alpha_{k1}}  \vec{p}_{{\rm Tx},\alpha_{k1}}
       &\ldots
       &\vec{p}_{{\rm Rx},1}^{~T} H_{1 \alpha_{kL_t}} \vec{p}_{{\rm Tx},\alpha_{kL_t}}\\
       \vdots  &\ddots  &\vdots \\
       \vec{p}_{{\rm Rx},N_{\rm r}}^{~T} H_{N_{\rm r} \alpha_{k1}} \vec{p}_{{\rm Tx},\alpha_{k1}}
       &\ldots
       &\vec{p}_{{\rm Rx},N_{\rm r}}^{~T} H_{N_{\rm r} \alpha_{kL_t}} \vec{p}_{{\rm Tx}, \alpha_{kL_t}}
     \end{bmatrix},{~~}\\
     \end{aligned}
\end{equation}
\setlength{\arraycolsep}{5pt}where $k \in \big\{1, 2, ..., \binom{N_t}{L_t} \big\}$ corresponds to the $k$-th row of the column index matrix $C_{\rm Index}$.
From (\ref{eq:C_sum}) -- (\ref{eq:Sum_Squared_SV}), the achievable PR-HS-MIMO channel capacity for $\widetilde{H}^{\rm eff}_k$ in (\ref{eq:AllpossibleHeff}), $C_{{\rm PR-HS}, k}$ is
\setlength{\arraycolsep}{0.14em}
\begin{eqnarray}
 \label{eq:C_PR-HS}
 C_{{\rm PR-HS}, k} &=& \sum _{i=1}^{ R_{ \widetilde{H}_k^{\rm eff} } } \log_2 \left( 1 + \frac{P_i}{{\sigma_{\rm n}}^2} {\sigma_i}^2 \right) \nonumber\\
 &\leq&R_{ \widetilde{H}_k^{\rm eff} } {~} \log_2 \left( \frac{1}{ R_{ \widetilde{H}_k^{\rm eff} } }
   \frac{\epsilon}{ {\sigma_{\rm n}}^2 }  \sum _{i=1}^{ R_{ \widetilde{H}_k^{\rm eff} } } {\sigma_i}^2 \right),
\end{eqnarray}
\setlength{\arraycolsep}{5pt}where the sum of squared singular values, $\sum _{i=1}^{ R_{ \widetilde{H}_k^{\rm eff}} } {\sigma_i}^2$ in (\ref{eq:C_PR-HS}) can be, analogous to Section \ref{sec:Pol_Pre-Post_Coding}, reformulated as
\setlength{\arraycolsep}{0.14em}
\begin{eqnarray}
 \label{eq:Sum_squared_SV_Lt}
 \gamma_{{\rm sum}, k} &=& \sum _{i=1}^{ R_{ \widetilde{H}_k^{\rm eff}} } {\gamma_i} =
   \sum _{i=1}^{ R_{ \widetilde{H}_k^{\rm eff}} } {\sigma_i}^2 \nonumber \\
   &=& {\rm Tr} \left( \widetilde{H}_k^{\rm eff} \widetilde{H}_k^{\rm eff \dag} \right)
     = \sum _{n,m} \left| \widetilde{h}_{k,nm}^{\rm eff} \right|^2.
\end{eqnarray}
\setlength{\arraycolsep}{5pt}The objective is selecting the column index $k$ that maximize PR-HS-MIMO channel capacity, $C_{{\rm PR-HS}, k}$. For further theoretical derivation,
\setlength{\arraycolsep}{0.14em}
\begin{eqnarray}
 \label{eq:heffheff*}
  &&\widetilde{H}_k^{\rm eff} \widetilde{H}_k^{\rm eff \dagger} = \\
  &&{~~~~~~~}\begin{bmatrix}
       &\vec{p}_{\rm Rx,1}^{~T}H^{\rm PD}_{\rm Rx,1}\vec{p}_{\rm Rx,1}
       &\ldots
       &\vec{p}_{\rm Rx,1}^{~T}H^{\rm Off}_{\rm Rx,1N_r}\vec{p}_{\rm Rx,N_r} \\
       &\vdots &\ddots &\vdots \\
       &\vec{p}_{\rm Rx,N_r}^{~T}H^{\rm Off}_{\rm Rx,N_r1},\vec{p}_{\rm Rx,1}
       &\ldots
       &\vec{p}_{\rm Rx,N_r}^{~T}H^{\rm PD}_{\rm Rx,N_r},\vec{p}_{\rm Rx,N_r}
  \end{bmatrix}, \nonumber
\end{eqnarray}
\setlength{\arraycolsep}{5pt}where 
\begin{equation}
 \label{eq:off diag HPD local}
 H^{\rm Off}_{{\rm Rx},ij} = \sum _{j=1}^{L_{\rm t}}H_{i\alpha_{ij}}\vec{p}_{{\rm Tx},\alpha_{ij}}\vec{p}_{{\rm Tx},\alpha_{ij}}^{{~}T}H_{j\alpha_{ij}}^{\dagger}
\end{equation}
is the polarization determinant matrix of $\emph{off-diagonal}$ components. The algorithm to maximize the diagonal components of the matrix is described in (\ref{eq:Opt_Rx_Pol_vector}) -- (\ref{eq:Opt_Rx_Pol_angle}).  The algorithm described in Section \ref{sec:Joint_Pol_Pre-Post_Coding} maximizes the sum of squared singular values of $\widetilde{H}_k^{\rm eff}$ presented in (\ref{eq:Sum_squared_SV_Lt}), which significantly improves the channel capacity.

We propose two schemes of PR-HS to efficiently achieve polarization reconfigurable antenna selection. One of the primary contributions of this paper is enhancing the PR-HS-MIMO channel capacity, $C_{{\rm PR-HS}, k}$ based on two proposed schemes in this section. Both schemes aim to determine the antenna index set that optimizes performance. We either find the set associated with $k$ which has the greatest $C_{{\rm PR-HS}, k}$ in (\ref{eq:C_PR-HS}); or we can maximize its upper bound via comparing the sum of squared singular values, $\gamma_{{\rm sum}, k} = \sum _{i=1}^{ R_{ \widetilde{H}_k^{\rm eff}} } {\gamma_i}$ in (\ref{eq:Sum_squared_SV_Lt}). The choice between these schemes is a  trade-off between accuracy and complexity/computation time. For instance, in the scenario where $N_t = 8$ as in our simulation, after performing joint polarization pre-post coding for each $\widetilde{H}_k^{\rm eff}$, the PR-HS-MIMO channel capacity, $C_{{\rm PR-HS}, k}$ can be directly used to determine the set of $L_t$ Tx antenna elements. Meanwhile, in massive MIMO scenarios, e.g., $N_t = 64$, as in a 5G and beyond-5G gNB antenna panel \cite{Roh_5GBeamforming_ComMag14, Nam_FD-MIMO, Nam_FD-MIMO_Feasibility_JSAC17, Kwon_Oh_VTC20Fall}, considering the sum of squared singular values, $\gamma_{{\rm sum}, k}$ for PR-HS at the Tx may be the better option, due to the reduced complexity.

\subsection{Scheme-1: Element-Wise Polarization Reconfiguration}
 \label{subsec:Option1}
Polarization pre-post coding in Section \ref{sec:Pol_Pre-Post_Coding} is accomplished with $\widetilde{H}_k^{\rm eff}$ for each $k \in \big\{1, 2, ..., \binom{N_t}{L_t} \big\}$, corresponding to the Tx antenna indices in $k$-th row of (\ref{eq:Column_Idx_matrix}), in element-wise polarization reconfiguration. Based on the estimated optimal polarization vectors for each $\widetilde{H}_k^{\rm eff}$, the best $k$ in terms of the greatest PR-HS-MIMO channel capacity, $C_{{\rm PR-HS}, k}$, is selected as $L_t$ Tx antenna elements. As aforementioned, in the scenario where the complexity crucially affects the overall system performance, the sum of squared singular values, $\gamma_{{\rm sum}, k} = \sum _{i=1}^{ R_{ \widetilde{H}_k^{\rm eff}} } {\gamma_i}$ in (\ref{eq:Sum_squared_SV_Lt}) may be considered to determine the $L_t$ antenna elements.
We denominate Scheme-1 as \emph{element-wise (EW)} polarization reconfiguration, since optimal polarization vectors of every $\widetilde{H}_k^{\rm eff}$ for $k \in \big\{1, 2, ..., \binom{N_t}{L_t} \big\}$ are estimated. The EW optimal Rx-polarization vectors are defined as
\begin{eqnarray}
 \label{eq:EW polarizationRx}
 \vec{p}_{{\rm Rx},i}^{{~}\rm EW} &=& \underset{\vec{p}_{{\rm Rx},i}}{\arg \max}{~}\vec{p}_{{\rm Rx},i}^{{~}T}H^{\rm EW}_{{\rm Rx},i}\vec{p}_{{\rm Rx},i} {~}, \\
 \label{eq:RxHpd}
 H^{\rm EW}_{{\rm Rx},i} &=& \sum _{j=1}^{L_{\rm t}}H_{i\alpha_{ij}}\vec{p}_{{\rm Tx},\alpha_{ij}}\vec{p}_{{\rm Tx},\alpha_{ij}}^{{~}T}H_{i\alpha_{ij}}^{\dagger} {~},
\end{eqnarray} where (\ref{eq:RxHpd}) is \textit{EW Rx-polarization-determinant matrix}.
In a similar manner, the optimal polarization vectors at the Tx are described as
\begin{eqnarray}
 \label{eq:EW polarizationTx}
  \vec{p}_{{\rm Tx},\alpha_{ij}}^{{~}\rm EW} &=& \underset{\vec{p}_{{\rm Tx},\alpha_{ij}}}{ \arg \max}{~}\vec{p}_{{\rm Tx},\alpha_{ij}}^{{~}T}H^{\rm EW}_{{\rm Tx},\alpha_{ij}}\vec{p}_{{\rm Tx},\alpha_{ij}}, \\
   \label{eq:TxHpd}
 H^{\rm EW}_{{\rm Tx},\alpha_{ij}} &=& \sum _{i=1}^{N_{\rm r}}H_{i\alpha_{ij}}^{\dagger}\vec{p}_{{\rm Rx},i}\vec{p}_{{\rm Rx},i}^{{~}T}H_{i\alpha_{ij}}
\end{eqnarray}
where (\ref{eq:TxHpd}) is the \textit{EW Tx-polarization-determinant matrix}. 
With (\ref{eq:EW polarizationRx}) and (\ref{eq:EW polarizationTx}), each $\widetilde{H}^{\rm eff}_k$ is tuned with corresponding EW Tx/Rx-polarization vectors. That is,
\setlength{\arraycolsep}{0.14em}
\begin{eqnarray}
 \label{eq:LocalHeff} 
&&\widetilde{H}^{\rm eff}_{ {\rm EW}, k }{}={} \\ 
&&{}\begin{bmatrix}
       (\vec{p}_{{\rm Rx},1}^{~ \rm EW})^{T} H_{1 \alpha_{k1}}  \vec{p}_{{\rm Tx},\alpha_{k1}}^{~ \rm EW}
       \ldots
       (\vec{p}_{{\rm Rx},1}^{~ \rm EW})^{T} H_{1 \alpha_{kL_t}} \vec{p}_{{\rm Tx},\alpha_{kL_t}}^{~\rm EW}\\
       \vdots ~~~~~~~~~~~~ \ddots ~~~~~~~~~~~~~~ \vdots \\
       (\vec{p}_{{\rm Rx},N_{\rm r}}^{~\rm EW})^{T} H_{N_{\rm r} \alpha_{k1}} \vec{p}_{{\rm Tx},\alpha_{k1}}^{~\rm EW}
       \ldots
       \vec{p}_{{\rm Rx},N_{\rm r}}^{ {\rm ~EW} ~T}  H_{N_{\rm r} \alpha_{kL_t}} \vec{p}_{{\rm Tx}, \alpha_{kL_t}}^{~\rm EW}
     \end{bmatrix}. \nonumber 
\end{eqnarray}
\setlength{\arraycolsep}{5pt}

Besides the EW polarization reconfiguration scheme, as aforementioned in this section, we also propose two metrics, PR-HS-MIMO channel capacity $C_{{\rm PR-HS}, k}^{\rm EW}$ and the sum of squared singular values $\gamma_{{\rm sum}, k}^{\rm EW}$ to determine the estimated best set of $L_t$ Tx antenna elements corresponding to $k$. The metrics are expressed as

\setlength{\arraycolsep}{0.14em}
\begin{eqnarray}
 \label{eq:C_PR-HS_EW}
 C_{{\rm PR-HS}, k}^{\rm EW} &=& \sum _{i=1}^{ R_{ \widetilde{H}_{{\rm EW}, k}^{\rm eff} } } \log_2 \left( 1 + \frac{P_i}{{\sigma_{\rm n}}^2} {\sigma_i}^2 \right),\\
 \label{eq:gamma_sum_PR-HS_EW}
 \gamma_{{\rm sum}, k}^{\rm EW} &=& \sum _{i=1}^{ R_{ \widetilde{H}_{{\rm EW}, k}^{\rm eff}} } {\gamma_i} =
   \sum _{i=1}^{ R_{ \widetilde{H}_{{\rm EW}, k}^{\rm eff}} } {\sigma_i}^2
   = \sum _{n,m} \left| \widetilde{h}_{k,nm}^{\rm eff} \right|^2. {~~~}
\end{eqnarray}
\setlength{\arraycolsep}{5pt}

\subsection{Scheme-2: Global Polarization Reconfiguration}
 \label{subsec:Option2}
This section  provides another PR-HS-MIMO spatial multiplexing scheme with successful performance but less complexity than the EW polarization reconfiguration scheme presented in Section \ref{subsec:Option1}. The scheme of global polarization reconfiguration first, estimates the optimal polarization vectors based on the proposed polarization pre-post coding in Section \ref{sec:Pol_Pre-Post_Coding} with the full $N_r \times N_t$ channel impulse response matrix regardless of $L_t$, the number of Tx antenna elements to be selected. To reduce the complexity of PR-HS processing, the global effective channel impulse response matrix $H^{\rm eff}$ rather than any partial channel impulse response matrix $\widetilde{H}_k^{\rm eff}$, is taken into account such that $L_t$ Tx antenna elements are selected to maximize the metric, PR-HS-MIMO channel capacity or the sum of squared singular values. Hence, the optimal Tx- and Rx-polarization vectors are aligned with (\ref{eq:Opt_Tx_Pol_vector}) -- (\ref{eq:Opt_Rx_Pol_angle}). For this reason, we denominate Scheme-2 as \textit{Global} polarization reconfiguration; for consistency, the optimal Tx- and Rx- polarization vectors of Scheme-2 are, respectively, expressed as 
\begin{eqnarray}
 \label{eq:GlobalRxPolarizationVectors}
 \vec{p}_{{\rm Tx},j}^{\rm ~G} = \vec{p}_{{\rm Tx},j}^{\rm ~opt}{~},{~~~}
 \vec{p}_{{\rm Rx},i}^{\rm ~G} = \vec{p}_{{\rm Rx},i}^{\rm ~opt} {~}.
\end{eqnarray}
\setlength{\arraycolsep}{0.14em}
\begin{eqnarray}
 \label{eq:Global_H_eff} 
{}&&{}\widetilde{H}^{\rm eff}_{ {\rm G}, k}{}={}\\ 
{}&&{}\begin{bmatrix}{}
       (\vec{p}_{{\rm Rx},1}^{~ \rm G})^{T} H_{1 \alpha_{k1}}  \vec{p}_{{\rm Tx},\alpha_{k1}}^{~ \rm G}
       \ldots
       (\vec{p}_{{\rm Rx},1}^{\rm ~G})^T H_{1 \alpha_{kL_t}} \vec{p}_{{\rm Tx},\alpha_{kL_t}}^{~\rm G} \\
       \vdots ~~~~~~~~~~~~ \ddots ~~~~~~~~~~~~~~ \vdots \\
       (\vec{p}_{{\rm Rx},N_{\rm r}}^{~\rm G})^T  H_{N_{\rm r} \alpha_{k1}} \vec{p}_{{\rm Tx},\alpha_{k1}}^{~\rm G}
       \cdot\cdot
       (\vec{p}_{{\rm Rx},N_{\rm r}}^{\rm ~G})^T   H_{N_{\rm r} \alpha_{kL_t}}  \vec{p}_{{\rm Tx}, \alpha_{kL_t}}^{~\rm G}
     \end{bmatrix} . \nonumber        
\end{eqnarray}
\setlength{\arraycolsep}{5pt}

In the same fashion as EW polarization reconfiguration, we consider either PR-HS-MIMO channel capacity $C_{{\rm PR-HS}, k}^{\rm G}$ or the sum of squared singular values $\gamma_{{\rm sum}, k}^{\rm G}$ to determine the estimated best set of $L_t$ Tx antenna elements corresponding to the index $k$, which are

\setlength{\arraycolsep}{0.14em}
\begin{eqnarray}
 \label{eq:C_PR-HS_G}
 C_{{\rm PR-HS}, k}^{\rm G} &=& \sum _{i=1}^{ R_{ \widetilde{H}_{{\rm G}, k}^{\rm eff} } } \log_2 \left( 1 + \frac{P_i}{{\sigma_{\rm n}}^2} {\sigma_i}^2 \right),\\
 \label{eq:gamma_sum_PR-HS_G}
 \gamma_{{\rm sum}, k}^{\rm G} &=& \sum _{i=1}^{ R_{ \widetilde{H}_{{\rm G}, k}^{\rm eff}} } {\gamma_i} =
   \sum _{i=1}^{ R_{ \widetilde{H}_{{\rm G}, k}^{\rm eff}} } {\sigma_i}^2
   = \sum _{n,m} \left| \widetilde{h}_{k,nm}^{\rm eff} \right|^2.
\end{eqnarray}
\setlength{\arraycolsep}{5pt}

\subsection{Comparison of Two Schemes}
 \label{sec:contrasting2options}
The upper bound of PR-HS-MIMO channel capacity for the considered $L_t$ Tx antenna elements corresponding to $k$ in (\ref{eq:C_PR-HS}) includes the associated sum of squared singular values, $\gamma_{{\rm sum}, k}^{\rm EW}$ in (\ref{eq:gamma_sum_PR-HS_EW}) or $\gamma_{{\rm sum}, k}^{\rm G}$ in (\ref{eq:gamma_sum_PR-HS_G}) for EW or global polarization reconfiguration schemes, respectively. Furthermore, the joint polarization pre-post coding is also accomplished based on this metric as elaborated in Section \ref{sec:Pol_Pre-Post_Coding}. Hence, it is worth comparing $\gamma_{{\rm sum}, k}^{\rm EW}$ with $\gamma_{{\rm sum}, k}^{\rm G}$ resulting from EW and global polarization reconfiguration schemes for the considered $k$ in Sections \ref{subsec:Option1} and \ref{subsec:Option2}, respectively.

$\gamma_{{\rm sum}, k}^{\rm EW}$ is the sum of squared singular values for $\widetilde{H}_{ {\rm EW}, k}^{\rm eff}$; equivalently, the sum of eigenvalues for $\widetilde{H}_{ {\rm EW}, k}^{\rm eff} \widetilde{H}_{{\rm EW}, k}^{ {\rm eff} {~}\dagger }$ that is described as
\setlength{\arraycolsep}{0.14em}
\begin{eqnarray}
 \label{eq:HlocalHlocalH}  
  {}&&{}\widetilde{H}_{ {\rm EW}, k}^{\rm eff} \widetilde{H}_{{\rm EW}, k}^{ {\rm eff} {~}\dagger }{}={}\\ 
  {}&&{}\begin{bmatrix}
       (\vec{p}_{\rm Rx,1}^{\rm ~EW})^{T}H^{\rm EW}_{\rm Rx,1}\vec{p}_{\rm Rx,1}^{\rm ~EW}~~~~
       \ldots
       ~~(\vec{p}_{\rm Rx,1}^{\rm ~EW})^{T}H^{\rm Off}_{\rm Rx,1N_r}\vec{p}_{\rm Rx,N_r}^{\rm ~EW} \\
       \vdots ~~~~~~~~~~~~~\ddots~~~~~~~~~~~~~~\vdots \\
       (\vec{p}_{\rm Rx,N_r}^{\rm ~EW})^{T}H^{\rm Off}_{\rm Rx,N_r1},\vec{p}_{\rm Rx,1}^{\rm ~EW}
       \ldots
       (\vec{p}_{\rm Rx,N_r}^{\rm ~EW})^{T}H^{\rm EW}_{\rm Rx,N_r},\vec{p}_{\rm Rx,N_r}^{\rm ~EW}
  \end{bmatrix} \nonumber  
\end{eqnarray}
\setlength{\arraycolsep}{5pt}
On the other hand, $\gamma_{{\rm sum}, k}^{\rm G}$ is the sum of eigenvalues for $\widetilde{H}_{ {\rm G}, k}^{\rm eff} \widetilde{H}_{{\rm G}, k}^{ {\rm eff} {~}\dagger }$ that is expressed as
\setlength{\arraycolsep}{0.14em}
\begin{eqnarray}
 \label{eq:HGlobalHGlobalH}  
  &&\widetilde{H}_{ {\rm G}, k}^{\rm eff} \widetilde{H}_{{\rm G}, k}^{ {\rm eff} {~}\dagger }{}={}\\ 
  &&\begin{bmatrix}
       (\vec{p}_{\rm Rx,1}^{\rm ~G})^{T}H^{\rm G}_{\rm Rx,1}\vec{p}_{\rm Rx,1}^{\rm ~G}~~~
       \ldots
       ~~(\vec{p}_{\rm Rx,1}^{\rm ~G})^{T}H^{\rm Off}_{\rm Rx,1N_r}\vec{p}_{\rm Rx,N_r}^{\rm ~G} \\
       \vdots ~~~~~~~~~~~~ \ddots ~~~~~~~~~~~~~~\vdots \\
       (\vec{p}_{\rm Rx,N_r}^{\rm ~G})^{T}H^{\rm Off}_{\rm Rx,N_r1},\vec{p}_{\rm Rx,1}^{\rm ~G}
       \ldots
       (\vec{p}_{\rm Rx,N_r}^{\rm ~G})^{T}H^{\rm G}_{\rm Rx,N_r},\vec{p}_{\rm Rx,N_r}^{\rm ~G}
  \end{bmatrix}, \nonumber
\end{eqnarray}
\setlength{\arraycolsep}{5pt}where the \emph{global Rx-polarization-determinant matrix}, $H^{\rm G}_{\rm Rx,i}$ is
\begin{eqnarray}
 H^{\rm G}_{\rm Rx,i}= \sum _{j=1}^{L_{\rm t}}H_{i\alpha_{ij}}\vec{p}_{{\rm Tx},\alpha_{ij}}\vec{p}_{{\rm Tx},\alpha_{ij}}^{{~}T}H_{i\alpha_{ij}}^{\dagger} {~}.
\end{eqnarray}
The diagonal components of (\ref{eq:HlocalHlocalH}) are maximized with EW polarization reconfiguration, since $\widetilde{H}^{\rm eff}_{{\rm EW}, k}$ is tuned for each $k$ according to its optimal polarization vectors, as described in (\ref{eq:EW polarizationRx}) -- (\ref{eq:TxHpd}). Hence, $\gamma_{{\rm sum}, k}^{\rm EW}$ is maximized. Meanwhile, polarization vectors in global polarization reconfiguration are optimized for the full channel impulse response matrix, $H^{\rm eff}$ regardless of $k$; therefore, diagonal components of (\ref{eq:HGlobalHGlobalH}) are not maximized for $L_t < N_t$. That is, those diagonal elements are aligned with (\ref{eq:Opt_Tx_Pol_vector}) -- (\ref{eq:Opt_Rx_Pol_angle}), which maximizes diagonal elements of $H^{\rm eff}(H^{\rm eff})^{\dagger}$. Nonetheless, global polarization reconfiguration also successfully increases the sum of squared singular values, $\gamma_{{\rm sum}, k}^{\rm G}$ for the considered $k$ with less complexity than EW polarization reconfiguration even when applied to each $\widetilde{H}^{\rm eff}_{{\rm G}, k}$. Further, as $L_t$ approaches $N_t$, $\widetilde{H}_{ {\rm EW}, k}^{\rm eff}$ approaches $\widetilde{H}_{ {\rm G}, k}^{\rm eff}$. Although EW polarization reconfiguration outperforms global polarization reconfiguration, the latter offers an advantage in the complexity and the corresponding computation time.

\section{Effect of PR-HS on the channel}
 \label{sec:Stat of Polarization}
\begin{figure}[ht]
  \centering
   \includegraphics[width=0.48\textwidth, height=4in]{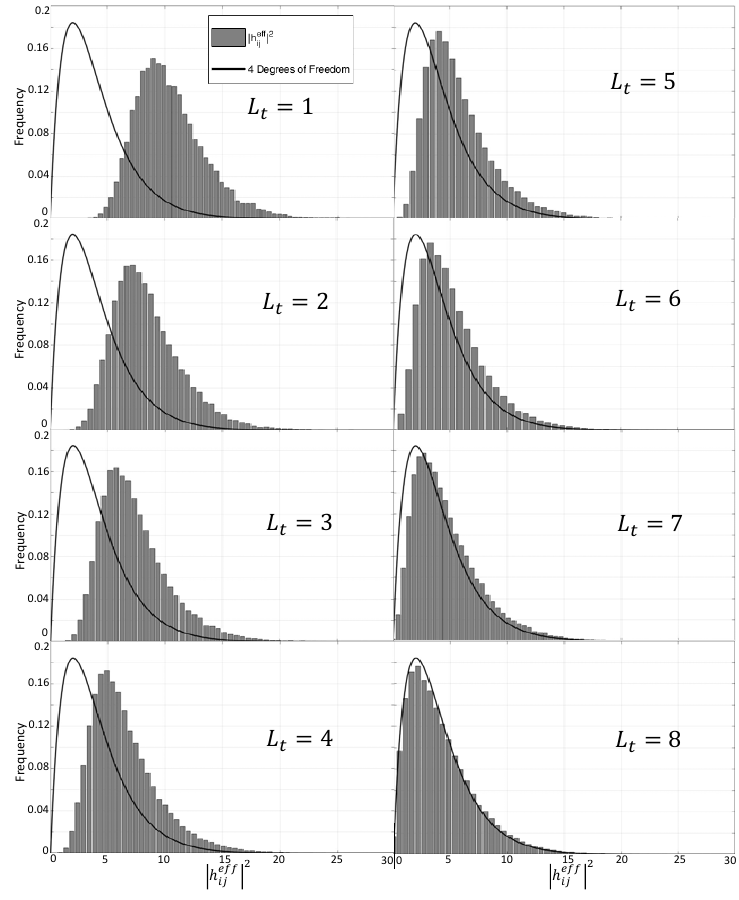}
  \caption{Empirical histogram of $|h^{\rm eff}_{ij}|^2$ in PR-HS; the comparison of chi-square distribution with 4 degrees of freedom.}
 \label{fig:chisquare}
\end{figure}

\begin{figure}[ht]
  \centering
   \includegraphics[width=0.474\textwidth]{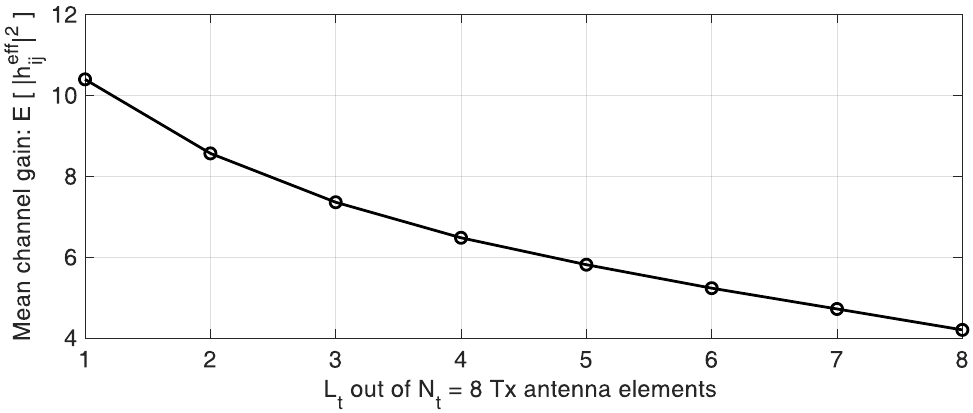}
  \caption{Impact of $L_t$ on the mean of effective channel gain in PR-HS-MIMO systems.}
 \label{fig:mean_EffChGain}
\end{figure}

It is worth emphasizing that the combination of polarization reconfiguration and hybrid antenna selection, i.e., PR-HS can provide significant improvement in effective channel gain, i.e., the squared envelop of $h^{\rm eff}_{ij}$.
The proposed polarization pre-post coding scheme is based on the closed-form derivation for the optimal polarization vectors at one end; whereas, the optimal polarization vectors at both ends of the Tx and Rx are achieved by the iterative methodology. For this reason, the complete analysis to reach the closed-form description is unfeasible. However, the comprehensive simulation results in Figs. \ref{fig:chisquare} -- \ref{fig:mean_EffChGain} show that the proposed PR-HS scheme substantially improves the system performance in terms of SNR, channel capacity and the associated distribution of channel gain.

Empirical distributions of channel gain, i.e., $|h^{\rm eff}_{ij}|^2$ are depicted in Fig. \ref{fig:chisquare}, where the PR-MIMO system applies hybrid antenna selection, i.e., PR-HS-MIMO with $N_t=8$ and $L_t \in \{1,2,..., 8 \}$. As $L_t$ decreases from $L_t = 8$ to $L_t = 1$, i.e., the number of residual antennas, $N_t - L_t$ increases, the distribution  of $|h^{\rm eff}_{ij}|^2$ and the corresponding average exhibit the significant improvement in channel gain.


As $L_t$ increases to reach $N_t=8$, the empirical histogram converge to chi-square distribution with 4 degrees of freedom. It is noteworthy that the channel gain, i.e., the squared envelope of the channel coefficient in conventional MIMO systems, has chi-square distribution with 2 degrees of freedom; whereas, PR-MIMO systems without hybrid antenna selection have 4 degrees of freedom. The reason is that PR-MIMO systems exploit the polarization domain where 2 degrees of freedom can be supported by orthogonal polarization, e.g., vertical and horizontal polarization.


Without loss of generality, an element in $H^{\rm eff}$ has the following description of its squared envelope.
\begin{eqnarray}
\label{eq:SISO}
|h^{\rm eff}_{11}|^2=\vec{p}_{{\rm Rx},1}^{~T} \big( H_{11}\vec{p}_{{\rm Tx},1}\vec{p}_{{\rm Tx},1}^{~T}H_{11}^{\dagger} \big)\vec{p}_{{\rm Rx},1}~.
\end{eqnarray}
From the analysis of (\ref{eq:SISO}), we can have an intuition for the mathematical interpretation of the aforementioned simulation results. However, owing to the page limit, Appendix A of \cite{Paul_Kwon_Molisch_Arxiv_AntSel} provides further analysis on (\ref{eq:SISO}).

\begin{figure}[ht]
  \centering
  \includegraphics[width=0.49\textwidth]{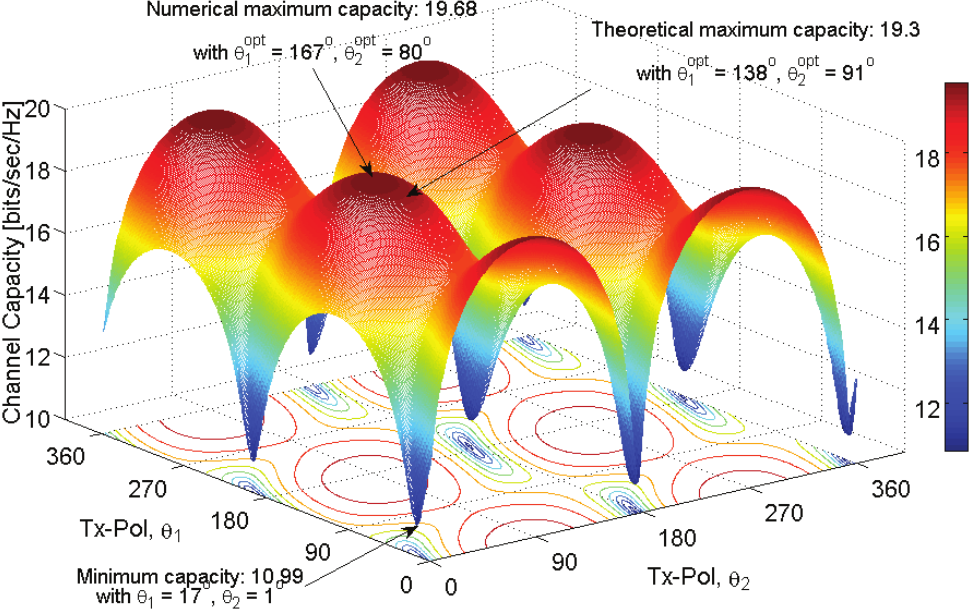}  
  \caption{Channel capacity for varying Tx-polarization; 30 dB SNR.}
  \label{fig:Capacity_SNR30dB}
\end{figure}

\section{Numerical Experiments and Results} \label{sec:Results}
\subsection{Simulations in PR-MIMO}
\begin{figure}[ht]
  \centering
  \includegraphics[width=0.48\textwidth]{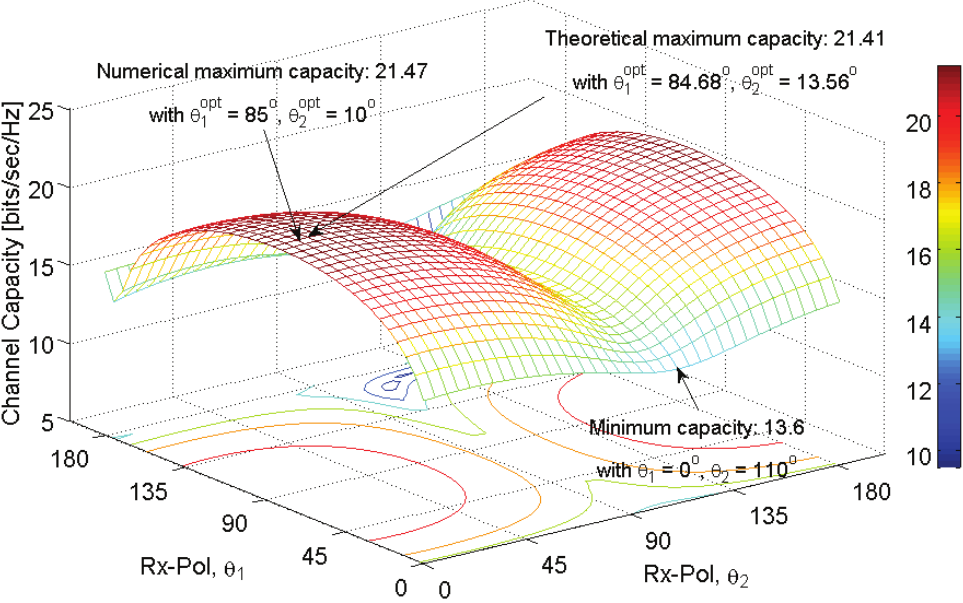}  
  \caption{Channel capacity for varying Rx-polarization with optimal Tx-polarization; 30 dB SNR.}
  \label{fig:Capacity_Pol_matching}
\end{figure}

\begin{figure}[ht]
  \centering
  \includegraphics[width=0.48\textwidth, height=2.5in]{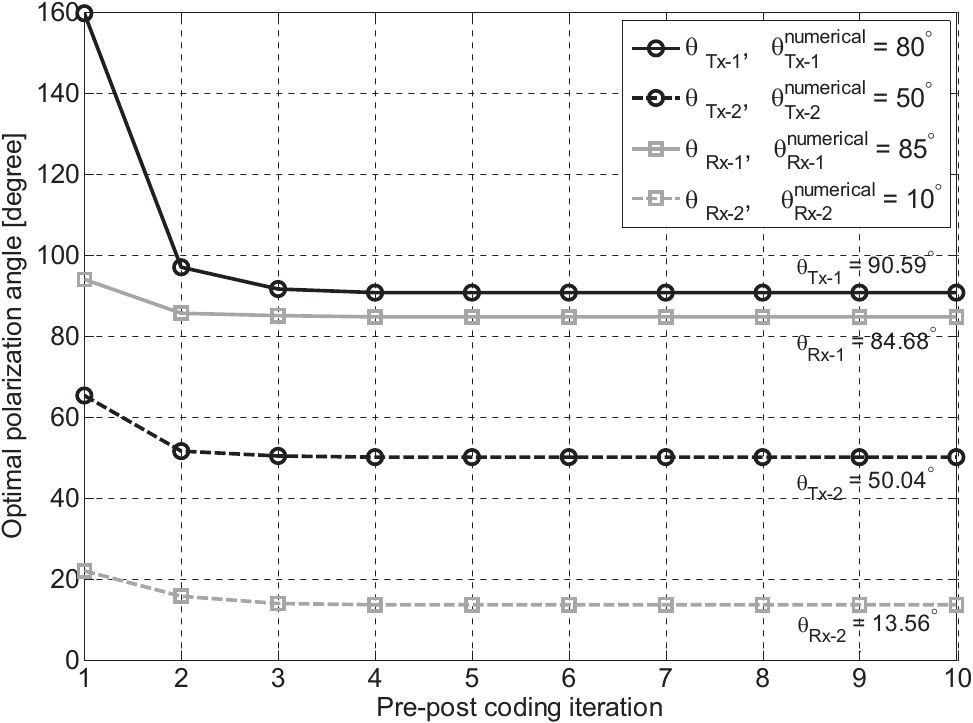}  
  \caption{Optimal Tx/Rx-polarization angles for the number of iterations; 30 dB.}
  \label{fig:Opt_Pol_vs_Iteration}
\end{figure}

\begin{figure}[ht]
  \centering
  \includegraphics[width=0.48\textwidth, height=2.5in]{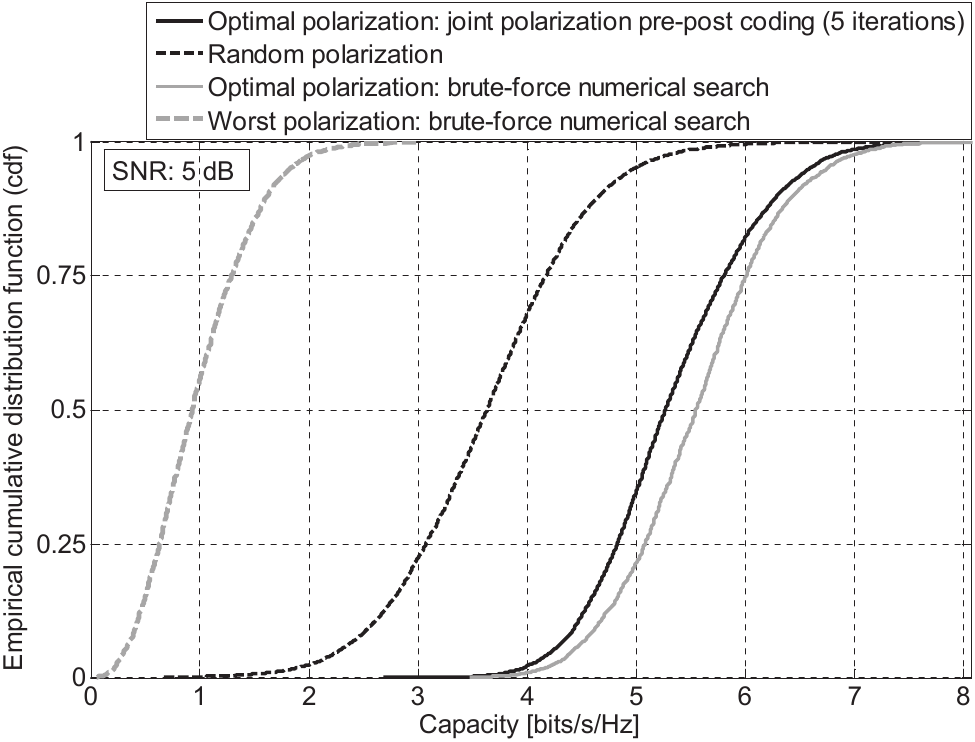}
  \caption{cdfs of channel capacity for the scenarios of joint polarization pre-post coding
           with five iterations, random Tx/Rx-polarization; optimal and the worst Tx/Rx-polarization
           through brute-force numerical search at 5 dB SNR.}
  \label{fig:cdf_Capacity_5dB}
\end{figure}

\begin{figure}[ht]
  \centering
  \includegraphics[width=0.48\textwidth, height=2.5in]{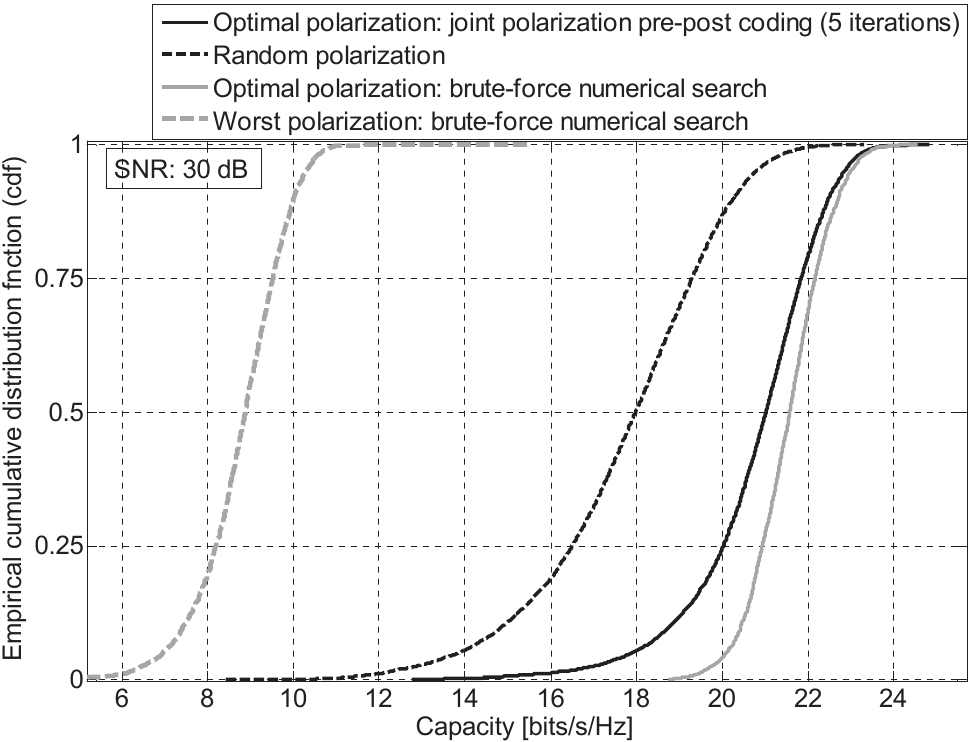}
  \caption{cdfs of channel capacity for the scenarios of joint polarization pre-post coding
           with five iterations, random Tx/Rx-polarization; optimal and the worst Tx/Rx-polarization
           through brute-force numerical search at 30 dB SNR.}
  \label{fig:cdf_Capacity_30dB}
\end{figure}

In this section we provide numerical evaluations of our schemes for optimum
polarization vectors and the associated PR-MIMO channel capacity. We also compare the results
to brute-force numerical optimization, where we step through all possible
(discretized) values of the Tx/Rx-polarization angle and choose the optimal
values that correspond to the maximum capacity for each PR-MIMO channel
realization.  The step width of the brute-force numerical search are,
respectively, $1^{\circ}$
in Fig. \ref{fig:Capacity_SNR30dB}; $5^{\circ}$ in Figs.
\ref{fig:Capacity_Pol_matching} and \ref{fig:Opt_Pol_vs_Iteration};
$10^{\circ}$ in Figs. \ref{fig:cdf_Capacity_5dB} and
\ref{fig:cdf_Capacity_30dB}.
Unless stated otherwise, we consider independent and identically distributed
(i.i.d.) Rayleigh fading channels with a cross-polarization discrimination,
${\rm XPD}=0 {\rm ~dB}$.

We first investigate the impact of polarization reconfiguration in several
deterministic channels. The PR-MIMO channel capacity in a $2 \times 2$ system with polarization reconfigurable antenna elements is shown for varying Tx-polarization angles in 
Fig. \ref{fig:Capacity_SNR30dB}, for 
the high SNR regime, i.e., 30 dB. 
The differences between theoretically and numerically obtained optimal Tx-polarization angles are considerable. This is due to the fact that the approximation (\ref{eq:Capacity_Jensen_Ineq}) is less accurate at higher SNRs.
However, the difference in capacity is still remarkably small. In case of polarization postcoding at the Rx, similar results are attained owing to the symmetry described in Sections \ref{sec:Pol_Precoding}; therefore, the results are omitted here. It is worth mentioning that optimal Tx- or
Rx-polarization vectors are not necessarily orthogonal, which corresponds to
$90^{\circ}$ difference in Tx- or Rx-polarization angles, as described
in Fig. \ref{fig:Capacity_SNR30dB}.
For the low SNR regime such as 5 dB SNR, the theoretically derived optimal Tx-polarization angles themselves have insignificant differences from numerically derived optimal Tx-polarization angles. The simulation results for the low SNR regime are omitted owing to the page limit.


The PR-MIMO channel capacity with the optimal Tx-polarization angles is
demonstrated to depend on varying Rx-polarization angles in Fig.
\ref{fig:Capacity_Pol_matching}, where we also consider the $2 \times 2$ PR-MIMO
system with polarization-agile antenna elements at both ends of the Tx and the
Rx.   The channel capacity exhibits substantial variation from 21.47
bits/sec/Hz to 13.60 bits/sec/Hz depending on Rx-polarization angles, although
Tx-polarization angles are set to the values that maximize channel capacity based on brute-force numerical search as will be shown in Fig. \ref{fig:Opt_Pol_vs_Iteration}, i.e.,
$\theta_{{\rm Tx}-1}=80^{\circ}$; $\theta_{{\rm Tx}-2}=50^{\circ}$.  This
result obviously shows that the polarization mismatching between the Tx and the
Rx can cause significant deterioration in the capacity of the whole system,
even if one end of the Tx and the Rx is already set to the optimal
polarization.
  However, the proposed scheme of joint polarization pre-post coding also shows
negligible difference from numerical results in both optimal Tx- and
Rx-polarization angles and channel capacity.

Locally optimal Tx- and Rx-polarization vectors at each iteration of joint
polarization pre-post coding are depicted in Fig.
\ref{fig:Opt_Pol_vs_Iteration} considering the same scenario of channel impulse
matrices and SNR as that in Fig. \ref{fig:Capacity_Pol_matching}. The $2 \times
2$ PR-MIMO system is again considered in this figure. Here, one iteration is a
sequential loop of polarization precoding and then polarization postcoding as
described in Section \ref{sec:Joint_Pol_Pre-Post_Coding}.
Tx/Rx-polarization vectors quickly converge; moreover, these global optimal
polarization vectors show relatively small difference from the numerical
optimum.

\begin{figure}[ht]
  \centering
  \includegraphics[width=0.485\textwidth]{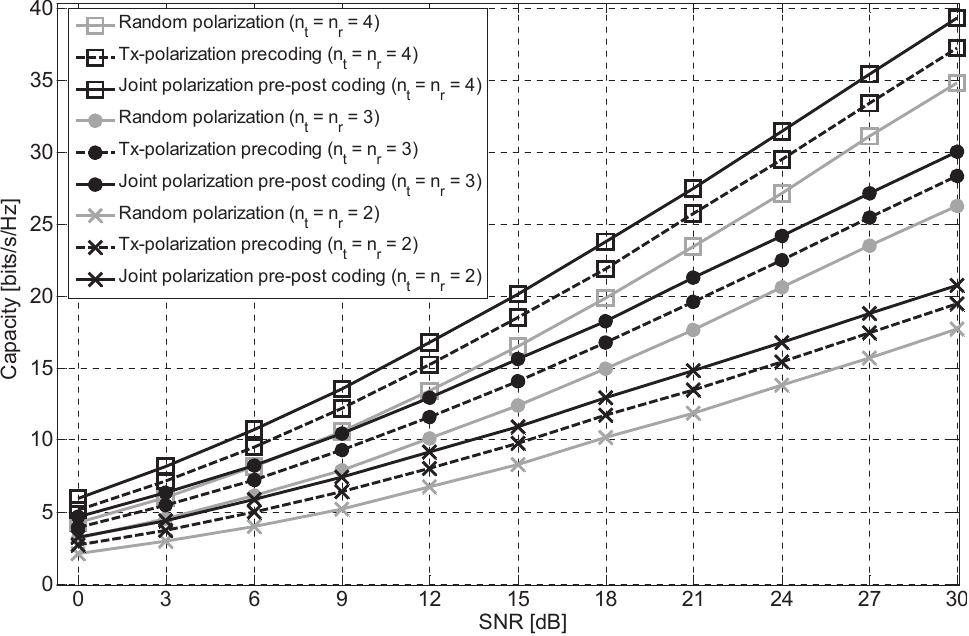}
  \caption{PR-MIMO channel capacity for the varying SNR in the scenarios of
           random Tx/Rx-polarization, Tx-polarization precoding, and joint polarization
           pre-post coding.}
  \label{fig:Capacity_vs_SNR}
\end{figure}


Figs.
\ref{fig:cdf_Capacity_5dB}~--~\ref{fig:Capacity_vs_SNR} investigate how much the joint polarization pre-post coding improves the
PR-MIMO channel capacity in a \textit{statistical} sense in .   The simulation
results are based on $10^4$ realizations of i.i.d. Rayleigh fading
channels. Figures \ref{fig:cdf_Capacity_5dB} and \ref{fig:cdf_Capacity_30dB}
depict cumulative density functions (cdfs) of the $2 \times 2$ PR-MIMO channel
capacity at 5 dB SNR and 30 dB SNR, respectively, for the scenarios of joint
polarization pre-post coding with Tx/Rx polarization-agile antennas; and random
Tx/Rx-polarization.  It is noteworthy that the cdf resulting from the fixed
Tx/Rx-polarization exhibits exactly the same cdf of random Tx/Rx-polarization
at each channel realization owing to the random generation of the i.i.d.
Rayleigh fading channels.  Furthermore, the cdfs of the optimal and the worst
Tx/Rx-polarization obtained by brute-force numerical search at each channel
realization, are presented in Figs. \ref{fig:cdf_Capacity_5dB} and
\ref{fig:cdf_Capacity_30dB}.  We perform five iterations for the joint
polarization pre-post coding at each PR-MIMO channel realization, which results
in the satisfactory convergence of Tx/Rx-polarization vectors on the global
optimal ones as demonstrated in Fig. \ref{fig:Opt_Pol_vs_Iteration}.

Our joint polarization pre-post coding significantly improves the PR-MIMO
channel capacity at both 5 dB SNR and 30 dB SNR.  In Fig.
\ref{fig:cdf_Capacity_5dB}, the probability of the PR-MIMO channel capacity less
than 4.2 bits/sec/Hz is 0.75 with 5 dB SNR and random Tx/Rx-polarization;
whereas, with the proposed joint polarization pre-post coding, the probability
of the capacity greater than 4.2 bits/sec/Hz is 0.94. For the improvement of
the PR-MIMO channel capacity at 30 dB SNR in Fig. \ref{fig:cdf_Capacity_30dB},
the probability of the capacity greater than 20 bits/sec/Hz is 0.75 in the
scenario of utilizing joint polarization pre-post coding, while that
probability is just 0.12 with random Tx/Rx-polarization. It is noteworthy that
the cdf curves of random Tx/Rx-polarization scenarios in both Figs.
\ref{fig:cdf_Capacity_5dB} and \ref{fig:cdf_Capacity_30dB} can be regarded as
the expectation of the cdf curves in a statistical sense when the joint
polarization pre-post coding is not utilized; however, the practical channel
capacity would exhibit substantial variations between the cdf curves associated
with the optimal and the worst Tx/Rx-polarization obtained by brute-force
numerical search in both figures.

\begin{figure*}[ht]
  \centering
  \includegraphics[width=1.00 \textwidth]{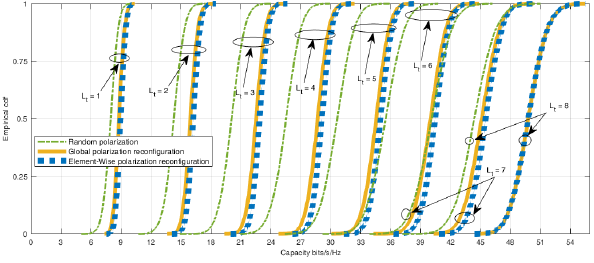}
  \caption{cdf curves of channel capacity at 20 dB for PR-HS-MIMO with element-wise (EW) and global polarization reconfiguration schemes, conventional HS-MIMO (Random Polarization); with a variety of $L_t$.}
  \label{fig:cdfLt1toLt8}
\end{figure*}

The joint polarization pre-post coding achieves significant improvement of
PR-MIMO channel capacity so that its cdf curve approaches the best
cdf curve obtained by brute-force numerical search, in particular, at 5 dB SNR
in Fig. \ref{fig:cdf_Capacity_5dB}. The cdf curve corresponding to the proposed
scheme also exhibits slight difference from the best cdf curve obtained from
brute-force numerical search at 30 dB SNR in Fig.
\ref{fig:cdf_Capacity_30dB}, compared to the significant difference between
that best-scenario cdf and the cdf of random Tx/Rx-polarization or, more
pronouncedly, between the best- and the worst-scenario cdfs obtained by the
numerical search.

Finally, we compare channel capacity for varying SNRs and varying number of
polarization-agile antennas in Fig. \ref{fig:Capacity_vs_SNR}.  For each
setting of the number of polarization-agile antennas, Fig.
\ref{fig:Capacity_vs_SNR} demonstrates three scenarios: joint polarization
pre-post coding at both link ends; Tx-polarization precoding only at the
transmitter; and random Tx/Rx-polarization as the baseline control.  In the
high SNR regime, utilizing our joint polarization pre-post coding improves PR-MIMO channel capacity with around 5 dB, 4 dB, and 3dB SNR gains in
the cases of $2 \times 2$, $3 \times 3$, and $4 \times 4$ PR-MIMO channels,
respectively.

Moreover, it is noteworthy that in a relatively low SNR regime below 9 dB, $2
\times 2$ and $3 \times 3$ PR-MIMO systems adopting the proposed joint
polarization pre-post coding accomplishes almost the same channel capacity as,
respectively, $3 \times 3$ and $4 \times 4$ MIMO systems with random
Tx/Rx-polarization.  We also note that the PR-MIMO system with our joint
polarization pre-post coding shows better channel capacity even than the PR-MIMO
system that has one more antenna element at both link ends of the transmitter
and receiver and uses random polarization, in the low SNR regime (below 3 dB). The degrees of freedom (slop at high SNR) are the same in all three cases, since they are determined by the number of antenna ports.

\subsection{Simulations in PR-HS-MIMO}
 \label{sec:ResultWAntennaselection}
In this section, we present simulation results focusing on the system performance when applying the proposed PR-HS-MIMO scheme. The results include improvement in channel capacity; and the comparison of EW and global polarization reconfiguration schemes in terms of channel capacity and selected Tx antenna indices. In the simulation, $N_t = 8$; $L_t = N_r \in \{1,2, ..., 8\}$; and we compare the PR-HS-MIMO channel capacity in (\ref{eq:C_PR-HS_EW}) and (\ref{eq:C_PR-HS_G}) rather than the sum of squared singular values in (\ref{eq:gamma_sum_PR-HS_EW}) and (\ref{eq:gamma_sum_PR-HS_G}) to select $L_t$ polarization reconfigurable Tx antenna indices. The reason is that for $N_t = 8$ total Tx antenna elements, directly comparing channel capacity has insignificant increase of the complexity beyond that of comparing the sum of squared singular values; whereas, the former provides the more appropriate hybrid selection of Tx antenna indices with the concomitant better improvement of channel capacity than the latter. 

Simulation results in Fig. \ref{fig:cdfLt1toLt8} show the impact of PR-HS-MIMO schemes on the channel capacity. Regardless of the number of selected Tx antenna elements, $L_t$, the proposed PR-HS-MIMO schemes outperform the conventional HS-MIMO scheme without polarization reconfiguration, i.e., the case of Random polarization in the legend of Fig. \ref{fig:cdfLt1toLt8}. The benefit of adopting polarization reconfigurable antenna elements becomes enlarged as $L_t$ increases.
Further, it is verified that EW polarization reconfiguration has better performance than global polarization reconfiguration although the difference in performance is not substantial in the scenario that $N_t = 8$. Hence, global polarization reconfiguration can also be a good suboptimal scheme considering the lower complexity and computation time than those of EW polarization reconfiguration. 

Examples of the selected Tx antenna indices in PR-HS-MIMO and conventional HS-MIMO, are given in Tables \ref{Table:Lt4} -- \ref{Table:Lt6}.
Even for a set of selected Tx antenna elements based on the conventional HS-MIMO, the channel capacity is improved via joint polarization pre-post coding after the hybrid antenna selection. However, the Tx antenna elements chosen by hybrid selection are different from the selection based on the proposed PR-HS-MIMO schemes. Selected antenna indices in EW polarization reconfiguration scheme are also different from those in global polarization reconfiguration scheme. 
Tables \ref{Table:Lt4} -- \ref{Table:Lt6} illustrate the selected Tx antenna indices in 10 independent channel realizations for $L_t = 4$ and $6$.

The conventional HS-MIMO scheme does not show full-matching of selected antenna indices with any of two PR-HS-MIMO schemes in the scenarios that $L_t = 4$ and $6$. On the other hand, the two PR-HS-MIMO schemes, i.e., EW and global polarization reconfiguration, have a considerable number of cases in which their selected antenna indices fully match as described in Tables \ref{Table:Lt4} -- \ref{Table:Lt6}.
It is worth emphasizing that estimation of optimal polarization vectors before the hybrid antenna selection stage is inevitable to have full benefit of joint polarization pre-post coding and the corresponding polarization reconfigurable antenna selection in PR-HS-MIMO spatial multiplexing.

\definecolor{Gray}{gray}{0.7}


\begin{table}[h!]
\centering
\begin{tabular}{|| c | c | c | c ||} 
 \hline
 Random & G & EW & Matching Index\\ [0.5ex] 
  \hline
 1     3     5     6 & \cellcolor{Gray}\textbf{2     4     5     7} & \cellcolor{Gray}\textbf{2     4     5     7} & 4 \\ 
  \hline
 2     3     7     8 & \cellcolor{Gray}\textbf{1     2     3     8} & \cellcolor{Gray}\textbf{1     2     3     8} & 4 \\
  \hline
 1     5     7     8 & \cellcolor{Gray}\textbf{1     5     7     8} & \cellcolor{Gray}\textbf{1     5     7     8} & 4 \\
  \hline
 2     3     6     7 & \textbf{1     3}     5     \textbf{6} & \textbf{1     3}     4     \textbf{6} & 3 \\
  \hline
 1     3     4     8 & \textbf{1}     5     \textbf{6     8} & \textbf{1}     3     \textbf{6     8} & 3 \\
  \hline
 1     3     5     7 & \textbf{2     3}     4     \textbf{7} & 1     \textbf{2     3     7} & 3 \\
  \hline
 1     2     4     8 & \cellcolor{Gray}\textbf{1     3     7     8} & \cellcolor{Gray}\textbf{1     3     7     8} & 4  \\
  \hline
 2     3     5     7 & \textbf{2}     4     \textbf{6     8} & \textbf{2}     6     \textbf{7     8} & 3 \\
  \hline
 1     2     3     4 & 2 \textbf{4     6     7} & \textbf{4     6     7}     8 & 3 \\
  \hline
 3     4     6     8 & \cellcolor{Gray}\textbf{4     5     7     8} & \cellcolor{Gray}\textbf{4     5     7     8} & 4 \\
 \hline 
\end{tabular}
\caption{Selected Tx Antenna Index when $L_t=4$}
\label{Table:Lt4}
\end{table}

\begin{table}[h!]
\centering
\begin{tabular}{|| c | c | c | c ||} 
 \hline
 Random & G & EW & Matching Index\\ [0.5ex] 
  \hline
 1     4     5     6     7     8 & \cellcolor{Gray}\textbf{1     4     5     6     7     8} & \cellcolor{Gray}\textbf{2     4     5     6     7     8} & 6 \\ 
  \hline
 1     3     4     5     7     8 & 1     \textbf{3     4     5     6}     7 & 2     \textbf{3     4     5     6}     8 & 4 \\
  \hline
 1     2     3     6     7     8 & 1     \textbf{2}     3     \textbf{4     7     8} & \textbf{2}     \textbf{4}     5     6     \textbf{7     8} & 4 \\
  \hline
 3     4     5     6     7     8 & \cellcolor{Gray}\textbf{1     2     4     5     7     8} & \cellcolor{Gray}\textbf{1     2     4     5     7     8} & 6 \\
  \hline
 3     4     5     6     7     8 & \textbf{2     4     5     6}     7     \textbf{8} & 1     \textbf{2     4     5     6     8} & 5 \\
  \hline
 1     2     3     5     6     8 & \textbf{2     3     5     6}     7     \textbf{8} & 1     \textbf{2     3     5     6     8} & 5 \\
  \hline
 2     3     4     5     7     8 & \cellcolor{Gray}\textbf{1     2     4     5     7     8} & \cellcolor{Gray}\textbf{1     2     4     5     7     8} & 6 \\
  \hline
 1     2     4     5     6     8 & \cellcolor{Gray}\textbf{1     3     4     6     7     8} & \cellcolor{Gray}\textbf{1     3     4     6     7     8} & 6 \\
  \hline
 1     3     4     6     7     8 & 2     \textbf{4}     5     \textbf{6     7     8} & 1     3     \textbf{4     6     7     8} & 4 \\
  \hline
 1     2     3     4     6     8 & \cellcolor{Gray}\textbf{1     3     4     5     6     8} & \cellcolor{Gray}\textbf{1     3     4     5     6     8} & 6 \\
 \hline 
\end{tabular}
\caption{Selected Tx Antenna Index when $L_t=6$}
\label{Table:Lt6}
\end{table}

\section{Conclusion} \label{sec:Conclusion}
In this paper, we proposed several novel schemes to support PR-HS-MIMO spatial multiplexing whose system is composed of multiple polarization reconfigurable antenna elements at both the Tx and Rx.  In the proposed iterative joint polarization pre-post coding, the local optimum usually reached the global optimum of Tx/Rx-polarization vectors within five iterations. The proposed method offers an energy-efficient as well as cost-efficient method to significantly increase channel capacity in PR-HS-MIMO spatial multiplexing. Furthermore, we proposed two PR-HS-MIMO schemes, i.e., EW and global polarization reconfiguration, and both schemes remarkably outperformed the conventional HS-MIMO without polarization reconfigurable antennas. The theoretical analysis along with extensive simulation results demonstrate the superior performance of the schemes proposed in this paper.

\section*{Acknowledgment}
Part of this work was supported financially by CSULB Foundation Fund, RS261-00181-10185.

\appendices 
\section{Elaboration of Effective Channel Gain}
This section elaborates the effective channel gain, i.e., the squared envelope of the effective channel coefficient resulting from the proposed polarization pre-post coding. From (\ref{eq:SISO}),
\setlength{\arraycolsep}{0.14em}
\begin{eqnarray}
 &&|h^{\rm eff}_{11}|^2=\vec{p}_{{\rm Rx},1}^{~T} \big( H_{11}\vec{p}_{{\rm Tx},1}\vec{p}_{{\rm Tx},j}^{~T}H_{11}^{\dagger} \big) 
 \vec{p}_{{\rm Rx},1} \\
 &&= \big[\cos(\theta_{{\rm Rx},1}) ~ \sin(\theta_{{\rm Rx},1})\big]
 \begin{bmatrix}
 		h^{\rm PD}_{{\rm Rx},11} &h^{\rm PD}_{{\rm Rx},12} \\ \nonumber
 		h^{\rm PD}_{{\rm Rx},21} &h^{\rm PD}_{{\rm Rx},22}
 \end{bmatrix}
  \begin{bmatrix}
 		\cos(\theta_{{\rm Rx},1}) \\ 
 		\sin(\theta_{{\rm Rx},1})		
 \end{bmatrix},
\end{eqnarray}  \label{eq:Heffsquared}
\setlength{\arraycolsep}{5pt}where 
\begin{eqnarray}
 \label{eq:hPD11}
 h^{\rm PD}_{{\rm Rx},11} &=& |h^{\rm vv}_{11}|^2\cos^2({\theta_{{\rm Tx},1}}) \\ \nonumber
 &&+h^{\rm vh}_{11}(h^{\rm vv}_{11})^{*}\sin(\theta_{{\rm Tx},1})\cos({\theta_{{\rm Tx},1}}) \\ \nonumber
 &&+ h^{\rm vv}_{11}(h^{\rm vh}_{11})^{*}\sin(\theta_{{\rm Tx},1})\cos({\theta_{{\rm Tx},1}}) \\ \nonumber
 &&+|h^{\rm vh}_{11}|^2\sin^2({\theta_{{\rm Tx},1}}) \\ 
 h^{\rm PD}_{{\rm Rx},12} &=& h^{\rm vv}_{11}(h^{\rm hv}_{11})^*\cos^2({\theta_{{\rm Tx},1}}) \\ \nonumber
 &&+h^{\rm vh}_{11}(h^{\rm hv}_{11})^{*}\sin(\theta_{{\rm Tx},1})\cos({\theta_{{\rm Tx},1}}) \\ \nonumber
 &&+ h^{\rm vv}_{11}(h^{\rm hh}_{11})^{*}\sin(\theta_{{\rm Tx},1})\cos({\theta_{{\rm Tx},1}}) ~~~~~~~~~~~~~~~~~~\\ \nonumber
 &&+h^{\rm vh}_{11}(h^{\rm hh}_{11})^*\sin^2({\theta_{{\rm Tx},1}})	\label{eq:hPD12} \\ 
 h^{\rm PD}_{{\rm Rx},21} &=& h^{\rm hv}_{11}(h^{\rm vv}_{11})^*\cos^2({\theta_{{\rm Tx},1}}) \\ \nonumber
 &&+h^{\rm hh}_{11}(h^{\rm vv}_{11})^{*}\sin(\theta_{{\rm Tx},1})\cos({\theta_{{\rm Tx},1}}) \\ \nonumber
 &&+ h^{\rm hv}_{11}(h^{\rm vh}_{11})^{*}\sin(\theta_{{\rm Tx},1})\cos({\theta_{{\rm Tx},1}}) \\ \nonumber 
 &&+h^{\rm hh}_{11}(h^{\rm vh}_{11})^*\sin^2({\theta_{{\rm Tx},1}})	\label{eq:hPD21} \\
 h^{\rm PD}_{{\rm Rx},22}&=& |h^{\rm hv}_{11}|^2\cos^2({\theta_{{\rm Tx},1}}) \\ \nonumber
 &&+h^{\rm hh}_{11}(h^{\rm hv}_{11})^{*}\sin(\theta_{{\rm Tx},1})\cos({\theta_{{\rm Tx},1}}) \\ \nonumber
 &&+ h^{\rm hv}_{11}(h^{\rm hh}_{11})^{*}\sin(\theta_{{\rm Tx},1})\cos({\theta_{{\rm Tx},1}}) \\ \nonumber 
 &&+|h^{\rm hh}_{11}|^2\sin^2({\theta_{{\rm Tx},1}})~. \label{eq:hPD22}
\end{eqnarray}
Further, utilizing trigonometric identities, they result as 
\begin{eqnarray}
\label{eq:heff2solved}
 |h^{\rm eff}_{11}|^2 
 &=& \dfrac{1}{4}\bigg\{ |h^{\rm vv}_{11}|^2+|h^{\rm vh}_{11}|^2+|h^{\rm hv}_{11}|^2+|h^{\rm hh}_{11}|^2 \\ \nonumber
 &&+ \cos{(2\theta_{{\rm Tx},1})}\big[ |h^{\rm vv}_{11}|^2-|h^{\rm vh}_{11}|^2 \\ \nonumber
 &&+|h^{\rm hv}_{11}|^2-|h^{\rm hh}_{11}|^2 \big]\\ \nonumber
 &&+ \sin{(2\theta_{{\rm Tx},1})}\big[ h^{\rm vh}_{11}(h^{\rm vv}_{11})^*+h^{\rm vv}_{11}(h^{\rm vh}_{11})^* \\ \nonumber
 &&+h^{\rm hh}_{11}(h^{\rm hv}_{11})^*+h^{\rm hv}_{11}(h^{\rm hh}_{11})^*\big] \bigg\} \\ \nonumber
 &&+\dfrac{1}{4}\cos{(2\theta_{{\rm Rx},1})}\bigg\{ |h^{\rm vv}_{11}|^2+|h^{\rm vh}_{11}|^2 \\ \nonumber
 &&+|h^{\rm hv}_{11}|^2+|h^{\rm hh}_{11}|^2 \\ \nonumber
 &&+ \cos{(2\theta_{{\rm Tx},1})}\big[ |h^{\rm vv}_{11}|^2-|h^{\rm vh}_{11}|^2 ~~~~~~~~~~~~~~~~~~~~~\\ \nonumber
 &&+|h^{\rm hv}_{11}|^2-|h^{\rm hh}_{11}|^2 \big] \\ \nonumber
 &&+ \sin{(2\theta_{{\rm Tx},1})}\big[ h^{\rm vh}_{11}(h^{\rm vv}_{11})^*+h^{\rm vv}_{11}(h^{\rm vh}_{11})^* \\ \nonumber
 &&+h^{\rm hh}_{11}(h^{\rm hv}_{11})^*+h^{\rm hv}_{11}(h^{\rm hh}_{11})^*\big] \bigg\} \\ \nonumber
 &&+\dfrac{1}{4}\sin{(2\theta_{{\rm Rx},1})}\bigg\{ h^{\rm vh}_{11}(h^{\rm vv}_{11})^*+h^{\rm vv}_{11}(h^{\rm vh}_{11})^* \\ \nonumber
 &&+h^{\rm hh}_{11}(h^{\rm hv}_{11})^*+h^{\rm hv}_{11}(h^{\rm hh}_{11})^* \\ \nonumber
 &&+\cos{(2\theta_{{\rm Tx},1})}\big[ h^{\rm vh}_{11}(h^{\rm vv}_{11})^*-h^{\rm vv}_{11}(h^{\rm vh}_{11})^* \\ \nonumber 
 &&+h^{\rm hh}_{11}(h^{\rm hv}_{11})^*-h^{\rm hv}_{11}(h^{\rm hh}_{11})^*\big] \\ \nonumber
 &&+\sin{(2\theta_{{\rm Tx},1})}\big[ h^{\rm hh}_{11}(h^{\rm vv}_{11})^*+h^{\rm hv}_{11}(h^{\rm vh}_{11})^* \\ \nonumber
 &&+h^{\rm vh}_{11}(h^{\rm hv}_{11})^*+h^{\rm hv}_{11}(h^{\rm hh}_{11})^*\big] \bigg\}~.
\end{eqnarray}
Optimal Tx/Rx-polarization angles are determined based on iterative joint polarization pre-post coding; therefore, the distribution of the channel gain is obtained and portrayed in Fig. \ref{fig:chisquare}, based on empirical simulations.




\bibliographystyle{IEEEtran}
\bibliography{PR-HS-MIMO-TWC-Rev15}

\end{document}